\documentclass[twocolumn]{jpsj3}
\usepackage{txfonts}
\usepackage{multicol}
\usepackage{times}
\usepackage{color}
\usepackage{bm}
\usepackage{subfigure}
\usepackage{here}
\usepackage{amsmath}

\title{
Electric Octupole Order in Bilayer Rashba System
}

\author{Takanori Hitomi\thanks{E-mail: hitomi@scphys.kyoto-u.ac.jp} and 
Youichi Yanase
}
\inst{
Department of Physics, Kyoto University, Kyoto 606-8502, Japan
}

\recdate{}

\abst{
The odd-parity multipole is an emergent degree of freedom, leading to spontaneous inversion symmetry breaking. 
The odd-parity multipole order may occur by forming staggered even-parity multipoles in a unit cell. 
We focus on a locally noncentrosymmetric bilayer Rashba system, and study an odd-parity electric octupole order 
caused by the antiferro stacking of local electric quadrupoles. 
Analyzing the forward scattering model, we show that the electric octupole order is stabilized by a  
layer-dependent Rashba spin-orbit coupling. 
The roles of the spin-orbit coupling are clarified on the basis of the analytic formula of multipole susceptibility. 
The spin texture allowed in the $D_{2d}$ point group symmetry and its magnetic response are revealed. 
Furthermore, we show that the parity-breaking quantum critical point appears in the magnetic field. 
The possible realization of the electric octupole order in bilayer high-$T_{\rm c}$ cuprate superconductors is discussed. 
}

\begin{document}
\maketitle

\renewcommand{\k}{{\bm k}}
\renewcommand{\r}{{\bm r}}
\newcommand{\dd}{{\bm d}}
\newcommand{\kk}{{\bm k'}}
\newcommand{\kkk}{{\bm k''}}
\newcommand{\q}{{\bm q}}
\newcommand{\Q}{{\bm Q}}
\newcommand{\e}{\varepsilon}
\newcommand{\ee}{e}
\newcommand{\s}{{\mit{\it \Sigma}}}
\newcommand{\J}{\mbox{\boldmath$J$}}
\newcommand{\vv}{\mbox{\boldmath$v$}}
\newcommand{\Jh}{J}
\newcommand{\LL}{\mbox{\boldmath$L$}}
\renewcommand{\SS}{\mbox{\boldmath$S$}}
\newcommand{\MM}{\mbox{\boldmath$M$}}
\newcommand{\g}{\mbox{\boldmath$g$}}
\newcommand{\HH}{\mbox{\boldmath$H$}}
\newcommand{\hh}{\mbox{\boldmath$h$}}
\newcommand{\Tc}{$T_{\rm c}$ }
\newcommand{\Tcf}{$T_{\rm c}$}
\newcommand{\Hc}{$H_{\rm c2}$ }
\newcommand{\Hcf}{$H_{\rm c2}$}
\newcommand{\etal}{{\it et al.}: }
\newcommand{\SRO}{Sr$_2$RuO$_4$ }
\newcommand{\SROf}{Sr$_2$RuO$_4$}
\newcommand{\px}{p_{x}}
\newcommand{\py}{p_{y}}

%%%%%%%%%%%%%%%%%%%%%%%%%%%%%%%%%%%%%%%%%%%%%%%%%%%%%%%%%%%%%%%%%%%
%%%%%%%%%%%%%%%%%%%%%%%%%%%%%%%%%%%%%%%%%%%%%%%%%%%%%%%%%%%%%%%%%%%
%%%%%%%%%%%%%%%%%%%%%%%%%%%%%%%%%%%%%%%%%%%%%%%%%%%%%%%%%%%%%%%%%%%
%%%%%%%%%%%%%%%%%%%%%%%%%%%%%%%%%%%%%%%%%%%%%%%%%%%%%%%%%%%%%%%%%%%
%%%%%%%%%%%%%%%%%%%%%%%%%%%%%%%%%%%%%%%%%%%%%%%%%%%%%%%%%%%%%%%%%%%
\section{Introduction}
Exotic quantum phases induced by local parity violation have been a subject of recent interest. 
A sublattice-dependent antisymmetric spin-orbit coupling (ASOC) arises from the local parity violation in crystal structures. 
Previous studies of superconductivity in multilayers\cite{Yoshida_PDW_1, Yoshida_PDW_2, Watanabe_PDW, Yoshida_PDW_3}, 
magnetic quadrupole order in zigzag chains\cite{Magnetic_Quadrupole, Hayami_Zigzag_Multipole}, 
toroidal order in a honeycomb lattice\cite{Hayami_Toroidal_1,Hayami_Toroidal_2}, 
and electric octupole (EO) order in bilayer systems\cite{Hitomi_EO} 
have revealed exotic quantum states of matter induced by the ASOC. 
On the basis of the generalized multipole expansion~\cite{Dubovik}, some of them are classified into odd-parity multipole states 
beyond the paradigm of even-parity multipole order studied in $d$- and $f$-electron systems\cite{Kusunose}. 
When the multipole moment is appropriately defined, ferroic odd-parity multipole order is accompanied by 
spontaneous global inversion symmetry breaking. 
The odd-parity toroidal order was demonstrated in LiCoPO$_4$~\cite{Aken,Spaldin}, and magnetic quadrupole order 
has recently been implied in Sr$_2$IrO$_4$~\cite{Zhao,Matteo}.

In this paper, we focus on the odd-parity EO state and clarify the thermodynamic stability 
in the bilayer Rashba system. 
Carrying out multipole expansion around the inversion center at the middle of bilayers, 
we identify the antiferro stacking of local electric quadrupoles as the EO state~\cite{Hitomi_EO}. 
The ferro stacking is naturally classified into the conventional even-parity electric quadrupole (EQ) state. 
In the same way, various odd-parity multipoles may be constructed by staggered even-parity multipoles 
in locally noncentrosymmetric systems\cite{Magnetic_Quadrupole, Hayami_Zigzag_Multipole,Hayami_Toroidal_1,Hayami_Toroidal_2}. 

Among the various mechanisms of multipole order, we consider the forward scattering 
that leads to the spontaneous deformation of the Fermi surface\cite{Yamase_dPI_1, Yamase_dPI_2, Khavkine_dPI, Kee_dPI}. 
This situation is relevant to the two-dimensional (2D) Hubbard model when the Fermi surface is 
in the vicinity of the van Hove singularity; renormalization group theories have shown 
the $d$-wave Pomeranchuk instability (dPI) \cite{Halboth_Metzner_dPI_RG_1, Halboth_Metzner_dPI_RG_2, 
Honerkamp_dPI_RG, Metzner_dPI_RG}. 
The dPI has been investigated in many theoretical works~\cite{Yamase_dPI_1, Halboth_Metzner_dPI_RG_1, Halboth_Metzner_dPI_RG_2,Yamase_dPI_2, Khavkine_dPI, Kee_dPI,Yamase-Katanin,Yamase2007,Yamase_Bilayer_1,Yamase_Bilayer_2} 
inspired by experimental reports 
on nematic order in bilayer high-$T_{\rm c}$ cuprate superconductors~\cite{Ando2002} and 
bilayer ruthenate Sr$_3$Ru$_2$O$_7$~\cite{Borzi,Stingl,Mackenzie,Lester}. 
On the basis of the symmetry argument, the dPI is regarded as an EQ order with $O_{x^{2} - y^{2}}$ symmetry. 
However, when the dPI order parameter is antiferroically ordered between bilayers, the fourfold rotation symmetry 
combined with the mirror reflection with respect to the $xy$-plane ($S_4$ symmetry) is preserved. 
Instead, the inversion symmetry is spontaneously broken. Then, the order parameter is identified as 
the EO moment with $T_{(x^{2} - y^{2}) \, z}$ symmetry in real space~\cite{Hitomi_EO}. 
The electric charge distribution actually shows 
asymmetry of the $(x^{2} - y^{2}) \, z$-type from the inversion center at the center of bilayers. 
The EO moment is viewed as magnetic a quadrupole moment~\cite{Fu_multipole} 
(or equivalently spin-nematicity~\cite{Hitomi_EO}) in momentum space.

In theoretical studies on the dPI in bilayer systems\cite{Yamase_Bilayer_1, Yamase_Bilayer_2}, 
the EO state is not thermodynamically stable in the weak forward scattering region.  
Then, the EQ state characterized by the ferro stacking of the dPI order parameter gains kinetic energy, 
and thus, it is stable. 
It was also shown that at $T=0$, the quantum critical point is hidden 
by the first-order quantum phase transition.~\cite{Kee_dPI,Yamase-Katanin,Yamase2007,Yamase_Bilayer_1}  
In these theories, the layer-dependent Rashba ASOC due to the bilayer structure has been neglected, 
although unusual properties of the EO state originate from the ASOC~\cite{Hitomi_EO,Maruyama_multilayer_SC}. 
In this paper we uncover the dramatic roles of the Rashba ASOC in the bilayer forward scattering model. 
{\it A thermodynamically stable EO state with parity-breaking quantum critical point} is demonstrated.

This paper is organized as follows.
In Sect. 2, we outline the EO order in the bilayer Rashba system 
by sketching Fermi surfaces.
In Sect. 3, we introduce the forward scattering model, 
taking into account the layer-dependent Rashba ASOC. 
In Sect. 4, the analytic form of the multipole susceptibility is obtained by adopting the random phase approximation. 
In Sect. 5, we shows the main conclusions. 
We show that the EO state is stabilized by the layer-dependent Rashba ASOC in a certain parameter range. 
We clarify the mechanism on the basis of multipole susceptibilities. 
We also demonstrate the $D_{2d}$ spin texture in the electronic structure of the EO state. 
In Sect. 6, we show the phase diagram revealing the octupole quantum critical point under a magnetic field. 
The asymmetric deformation of the Fermi surface due to the in-plane magnetic field is also demonstrated. 
Finally, we give a brief summary, and discuss possible realization in high-$T_{\rm c}$ cuprate superconductors in Sect.~7.

\section{Sketch of Electric Quadrupole and Octupole States}

\begin{figure}[htbp]
  \begin{center}
    \includegraphics[width=0.4\hsize]{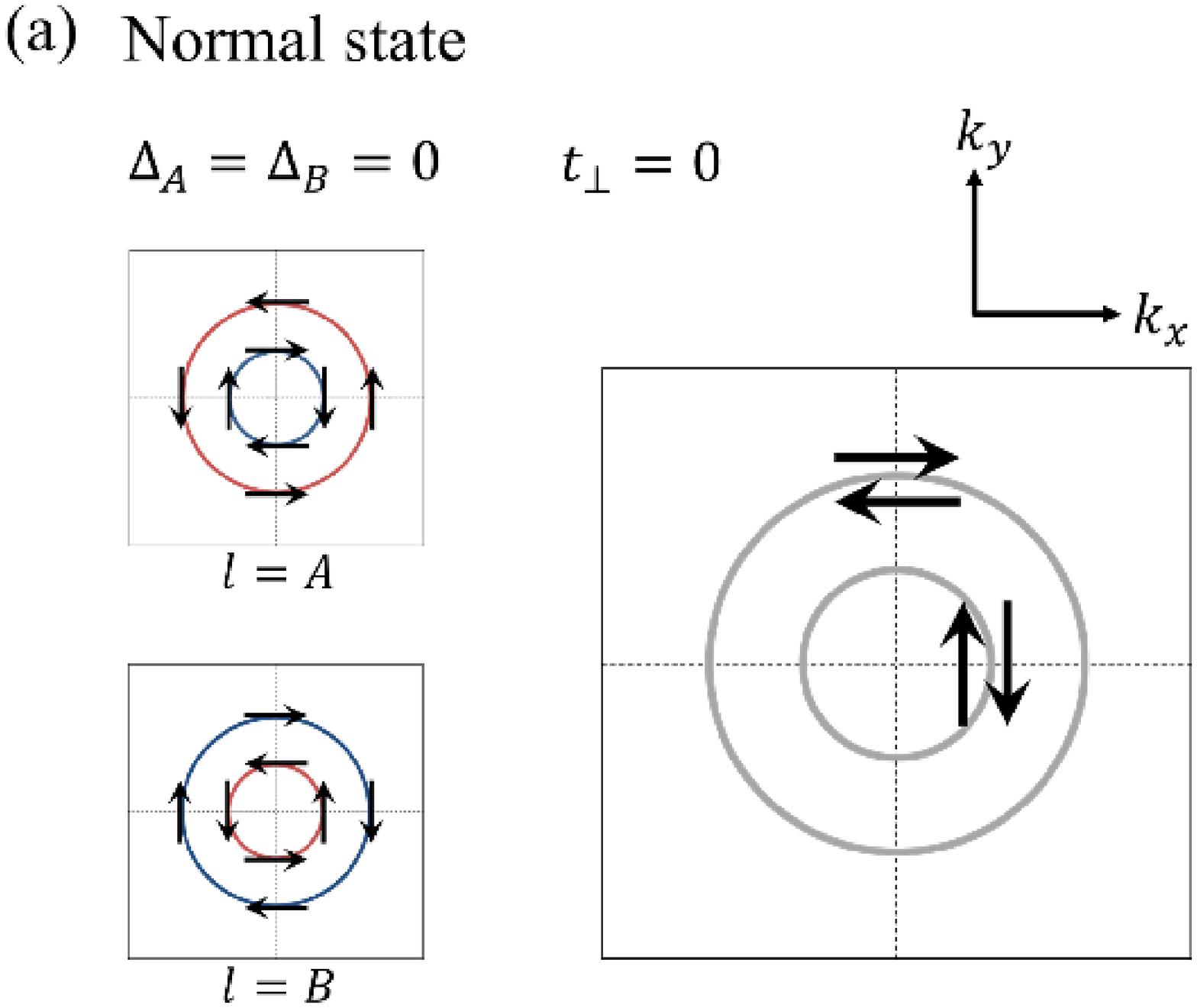}
%  \end{center}
%\end{figure}
%\begin{figure}[H]
%  \begin{center}
    \includegraphics[width=0.4\hsize]{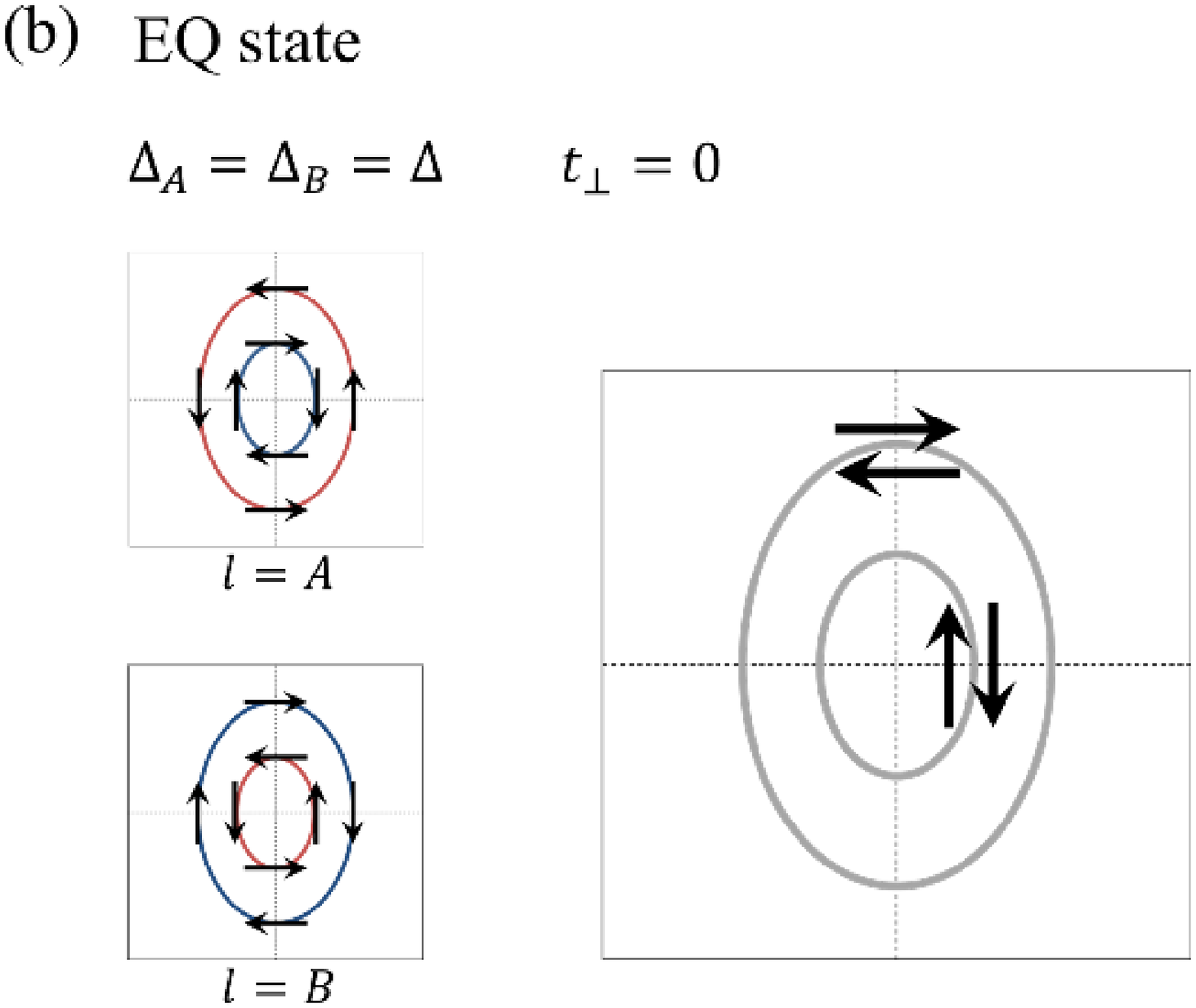}
    \includegraphics[width=0.4\hsize]{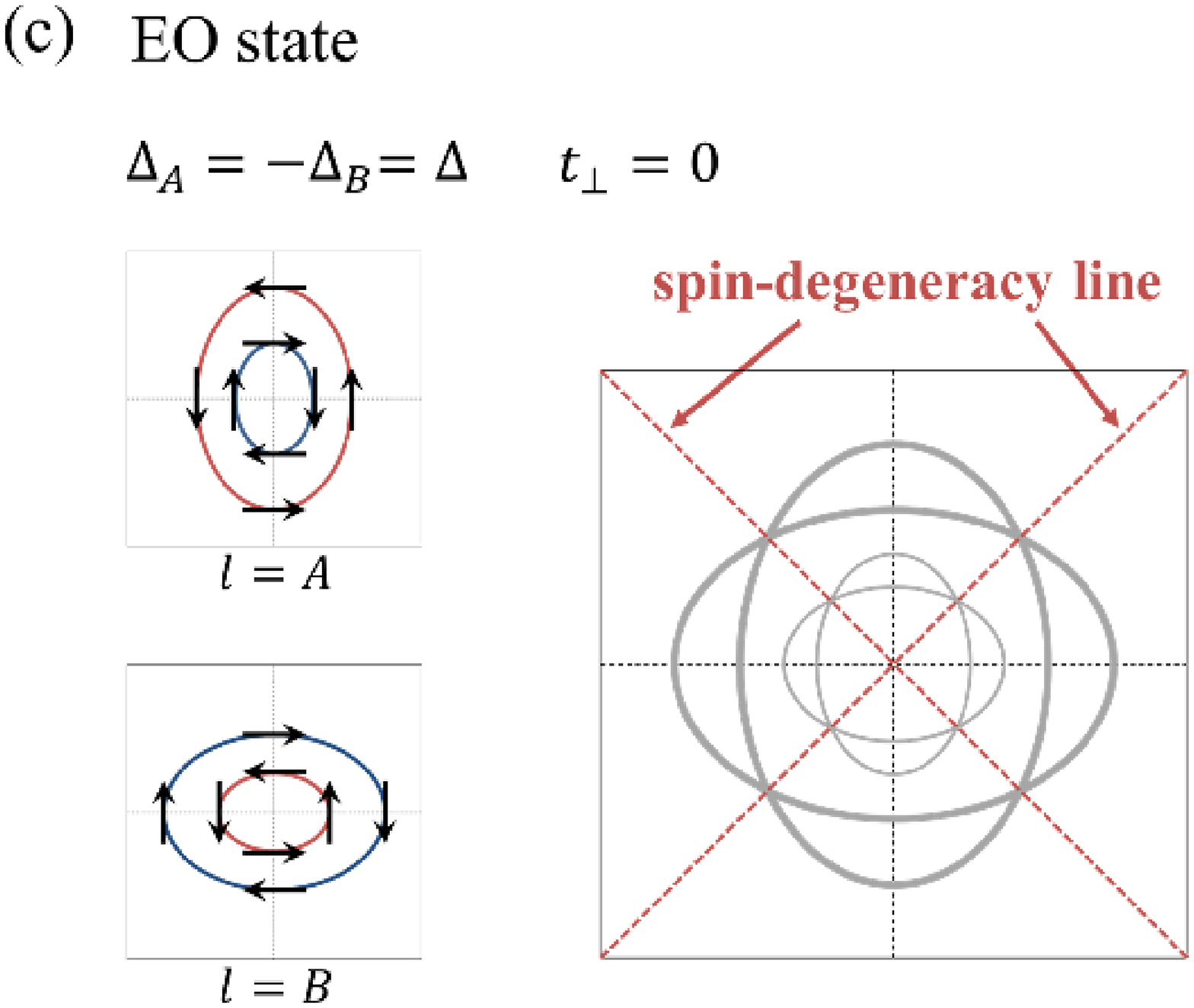}
    \includegraphics[width=0.4\hsize]{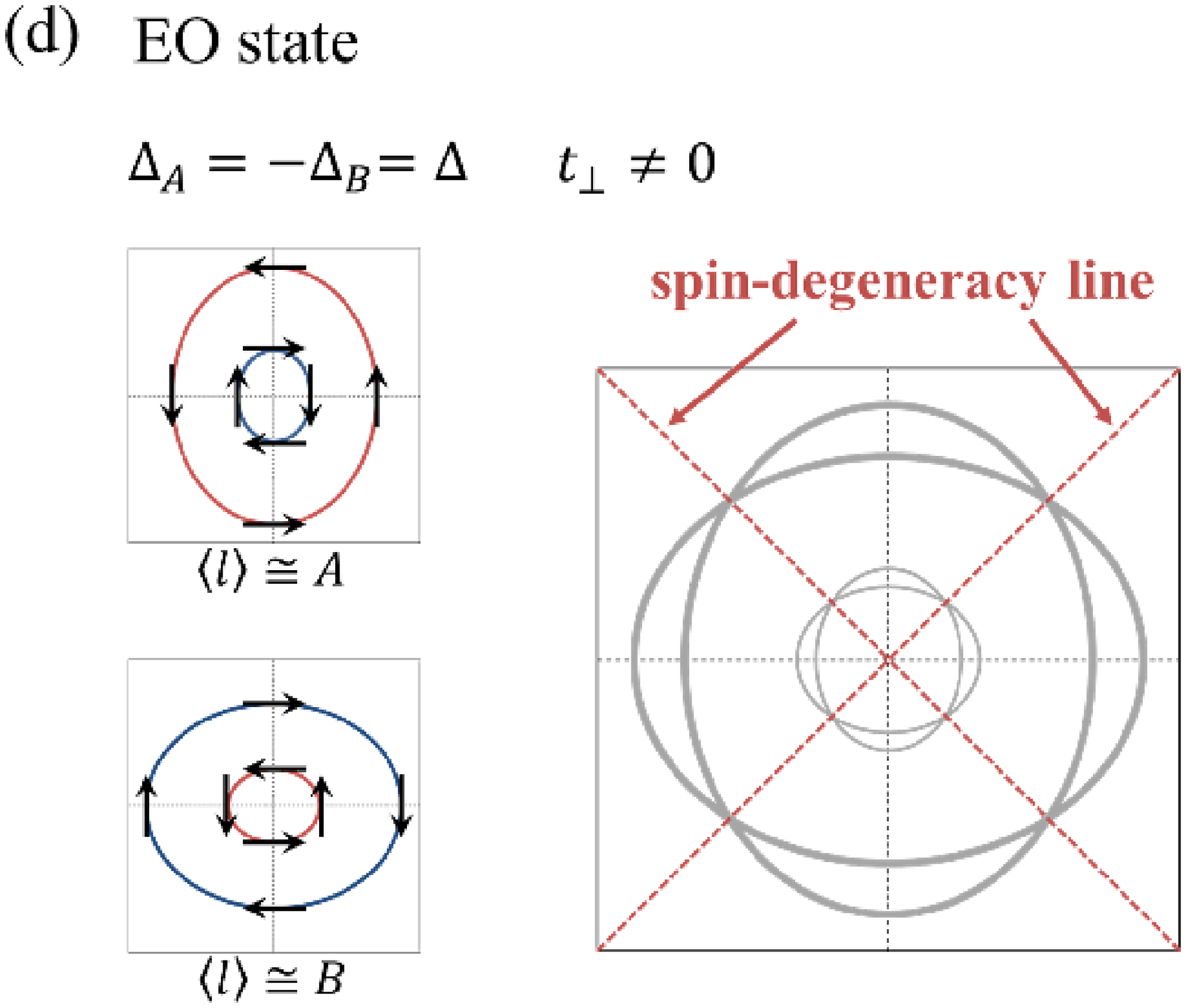}
    \caption{(Color online) Schematic of Fermi surfaces in bilayer Rashba system. 
(a) Normal state ($\Delta_{A} = \Delta_{B} = 0$), (b) EQ state ($\Delta_{A} = \Delta_{B} = \Delta$), 
and (c) EO state ($\Delta_{A} = -\Delta_{B} = \Delta$). The interlayer hopping is assumed to be zero, $t_{\perp} = 0$. 
The definition of $t_{\perp}$ and $\Delta_{l}$ is given in Sect.~3. 
The left panels show the Fermi surfaces in the layers $A$ and $B$, while the right panels show the total Fermi surfaces. 
The Fermi surfaces in each layer are split owing to the ASOC. 
The black arrows indicate the spin texture.
In the EQ state, the twofold degeneracy is preserved, but the  $C_4$ rotation symmetry is broken. 
On the other hand, owing to the spontaneous inversion symmetry breaking, the splitting of total Fermi surfaces 
occurs in the EO state, although the $C_4$ rotation symmetry of the Fermi surfaces is preserved. 
In (d), we sketch the Fermi surfaces in the EO state with a finite interlayer hopping  $t_{\perp} \ne 0$. 
Qualitative features are not altered. 
    }
    \label{fig:FSs_each_multipole_state}
  \end{center}
\end{figure}

Before conducting theoretical analysis, we sketch the electric multipole order in the bilayer Rashba system. 
For clarity, we neglect the interlayer hopping $t_\perp$ in this section. 
Then, the two layers are completely decoupled; thus, the Fermi surfaces are decomposed 
in terms of the layers as illustrated in Fig.~\ref{fig:FSs_each_multipole_state}.

First, Fig.~\ref{fig:FSs_each_multipole_state}(a) shows the 2D Fermi surfaces in the normal state. 
In both layers, the Rashba ASOC induces the spin splitting in the Fermi surface. 
Although the shapes of the Fermi surfaces are equivalent between layers, the spin textures are opposite because of the 
opposite signs of the Rashba ASOC. Thus, Kramers pairs are formed between the layers~\cite{Maruyama_multilayer_SC,Fischer}. 
The twofold degeneracy in the band structure is preserved in agreement with the global inversion symmetry. 
We emphasize that the twofold degeneracy comes from the composite degree of freedom composed of spin and sublattice. 
Recently, such ``hidden spin splitting'' has been demonstrated in various locally noncentrosymmetric compounds\cite{Zhang2014,Riley2014,Klein2016}.

When the ferroic dPI occurs, Fermi surfaces are deformed as in Fig.~\ref{fig:FSs_each_multipole_state}(b). 
Since the deformation equivalently occurs in the two layers, Kramers degeneracy in the band structure is preserved. 
The rotation symmetry is reduced from $C_{4}$ to $C_{2}$, and thus, this state is regarded as the EQ state or 
the nematic state~\cite{Yamase_dPI_1, Halboth_Metzner_dPI_RG_1, Halboth_Metzner_dPI_RG_2,Yamase_dPI_2, Khavkine_dPI, Kee_dPI,Yamase-Katanin,Yamase2007}.

On the other hand, the order parameters of dPI are opposite between layers in the EO state. 
Then, the Fermi surfaces of each layer are deformed oppositely, as shown 
in the left panel of Fig.~\ref{fig:FSs_each_multipole_state}(c). 
The $C_{4}$ symmetry of Fermi surfaces remains to be preserved 
[right panel of Fig.~\ref{fig:FSs_each_multipole_state}(c)]. 
However, the twofold degeneracy is lifted as a consequence of the spontaneous inversion symmetry breaking. 
The layer-dependent Rashba ASOC plays a vital role in the spin splitting in total Fermi surfaces. 
In this way, the sublattice-dependent ASOC gives rise to unconventional properties 
in the odd-parity multipole state~\cite{Magnetic_Quadrupole,Hayami_Zigzag_Multipole,Hayami_Toroidal_1,Hayami_Toroidal_2,Hitomi_EO}.

%%%%%%%%%%%%%%%%%%%%%%%%%%%%%%%%%%%%%%%%%%%%%%%%%%%%%%%%%%%%%%%%%%%
%%%%%%%%%%%%%%%%%%%%%%%%%%%%%%%%%%%%%%%%%%%%%%%%%%%%%%%%%%%%%%%%%%%
%%%%%%%%%%%%%%%%%%%%%%%%%%%%%%%%%%%%%%%%%%%%%%%%%%%%%%%%%%%%%%%%%%%
%%%%%%%%%%%%%%%%%%%%%%%%%%%%%%%%%%%%%%%%%%%%%%%%%%%%%%%%%%%%%%%%%%%
%%%%%%%%%%%%%%%%%%%%%%%%%%%%%%%%%%%%%%%%%%%%%%%%%%%%%%%%%%%%%%%%%%%

Note that the spin degeneracy is preserved along lines $|k_{x}| = |k_{y}|$, as indicated by the red dashed lines 
in  Fig.~\ref{fig:FSs_each_multipole_state}(c). This degeneracy arises from the $d$-wave form of 
the Pomeranchuk instability, and it is not protected by symmetry.  
Indeed, this accidental degeneracy is lifted in a generic EO state,  
for instance, by orbital polarization\cite{Hitomi_EO}. 

Finally, the role of the interlayer hopping $t_{\perp} \ne 0$ is discussed. 
Owing to the interlayer hopping, the level repulsion increases the energy difference between bonding and anti-bonding bands. 
However, the spin splitting in each Fermi surface is suppressed because the interlayer hopping competes with the layer-dependent Rashba ASOC~\cite{Maruyama_multilayer_SC}. 
Despite these quantitative differences, the qualitative features of Fermi surfaces are not altered by introducing the interlayer hopping. 
%In other words, the electronic structure in the odd-parity EO state is captured by neglecting the interlayer hopping. 

%%%%%%%%%%%%%%%%%%%%%%%%%%%%%%%%%%%%%%%%%%%%%%%%%%%%%%%%%%%%%%%%%%%
%%%%%%%%%%%%%%%%%%%%%%%%%%%%%%%%%%%%%%%%%%%%%%%%%%%%%%%%%%%%%%%%%%%
%%%%%%%%%%%%%%%%%%%%%%%%%%%%%%%%%%%%%%%%%%%%%%%%%%%%%%%%%%%%%%%%%%%
%%%%%%%%%%%%%%%%%%%%%%%%%%%%%%%%%%%%%%%%%%%%%%%%%%%%%%%%%%%%%%%%%%%
%%%%%%%%%%%%%%%%%%%%%%%%%%%%%%%%%%%%%%%%%%%%%%%%%%%%%%%%%%%%%%%%%%%
\section{Formulation} 
%%%%%%%%%%%%%%%%%%%%%%%%%%%%%%%%%%%%%%%%%%%%%%%%%%%%%%%%%%%%%%%%%%%
%%%%%%%%%%%%%%%%%%%%%%%%%%%%%%%%%%%%%%%%%%%%%%%%%%%%%%%%%%%%%%%%%%%
%%%%%%%%%%%%%%%%%%%%%%%%%%%%%%%%%%%%%%%%%%%%%%%%%%%%%%%%%%%%%%%%%%%
\subsection{Model}
To investigate multipole order in the bilayer Rashba system, 
we analyze the forward scattering model given by 
\begin{align}
H &= H_{\rm{kin}} + H_{\rm{ASOC}} + H_{\perp} + H_{\rm{f}} ,   \label{eq:1} \\
H_{\rm{kin}}   &= \sum_{\bm{k}} \sum_{s=\uparrow, \downarrow} \sum_{l=A,B} \varepsilon_{\bm{k}} \, c^{\dagger}_{\bm{k}sl} \, c_{\bm{k}sl} ,  \label{eq:2} \\
H_{\rm{ASOC}}  &= \sum_{\bm{k},s,s',l} \alpha_{l} \, \bm{g}_{\bm{k}} \cdot \bm{\sigma}^{ss'} c^{\dagger}_{\bm{k}sl} \, c_{\bm{k}s'l} ,   \label{eq:3} \\
H_{\perp}     &= t_{\perp} \sum_{\bm{k},s} [c^{\dagger}_{\bm{k}sA} c_{\bm{k}sB} + \rm{h.c.}] ,           \label{eq:4} \\ 
H_{\rm{f}}    &= - \frac{g_{1}}{2N} \sum_{\bm{k},\bm{k}',l} d_{\bm{k}} d_{\bm{k}'} n_{\bm{k}l} n_{\bm{k}'l} \notag \\
             &  \qquad - \frac{g_{2}}{2N} \sum_{\bm{k},\bm{k}'} d_{\bm{k}} d_{\bm{k}'} [n_{\bm{k}A} n_{\bm{k}'B} + n_{\bm{k}B} n_{\bm{k}'A}], \label{eq:5}
\end{align}
where $c_{\bm{k}sl}$ ($c^{\dagger}_{\bm{k}sl}$) is the annihilation (creation) operator of an electron with spin $s=\uparrow,\downarrow$ and the wave vector $\bm{k}$ on the layer $l=A,B$, and $n_{\bm{k}l} = \sum_{s} c^{\dagger}_{\bm{k}sl} c_{\bm{k}sl}$ is the number operator.
$\bm{\sigma} = (\sigma_{x}, \sigma_{y}, \sigma_{z})$ denotes Pauli matrices and $N$ is the number of sites per layer. 

The first term $H_{\rm{kin}}$ is the kinetic energy term and $\varepsilon_{\bm{k}}$ is the 2D dispersion in a square lattice, i.e., $\varepsilon_{\bm{k}} = -2t_{1}(\cos k_{x} + \cos k_{y}) - 4t_{2} \cos k_{x} \cos k_{y} - \mu$. 
The chemical potential $\mu$ is included in the dispersion relation.
For the nearest- and next-nearest-neighbour hopping amplitude, 
we choose $t_{1} = 1$ as the unit of energy and assume $t_{2} = 0.35$. 

The second term $H_{\rm{ASOC}}$ represents the layer-dependent Rashba ASOC arising from the local violation of 
inversion symmetry at each atomic site. 
It has been shown that the Rashba ASOC is induced by the combination of the atomic LS coupling and 
local parity mixing in electron functions~\cite{Yanase_ASOC,Zhong-Held,Yanase_multi_orbital_SOC}. 
Thus, the origin of Rashba ASOC does not require global inversion symmetry breaking, 
and indeed local violation of inversion symmetry is a sufficient condition for the presence of local Rashba ASOC. 
In contrast to globally noncentrosymmetric systems, the Rashba ASOC is sublattice-dependent and the average 
in the unit cell disappears in locally noncentrosymmetric systems\cite{Maruyama_multilayer_SC,Fischer,Zhang2014,Riley2014}. 
In bilayers, the coupling constant is $(\alpha_{A} ,\alpha_{B}) = (\alpha, -\alpha)$. 
In the present work, for simplicity, the Rashba ASOC is characterized by a g-vector of simple form, 
$\bm{g}_{\bm{k}} = (-\sin k_{y} ,\sin k_{x}, 0)$. 
This ``hidden'' spin-orbit coupling may stabilize the EO state as we show later.

The third term $H_{\perp}$ represents the interlayer hopping. 
Considering quasi-2D bilayer systems, we assume a small interlayer hopping amplitude, $t_\perp=0.1$. 
It has been shown that the ratio of the ASOC to the interlayer hopping, i.e., $\alpha / t_{\perp}$, offers 
a measure of the effect of the ASOC \cite{Maruyama_multilayer_SC}. Therefore, unusual properties arising from 
the spin-orbit coupling are enhanced in quasi-2D systems.

Finally, the forward scattering term $H_{\rm{f}}$ is introduced. 
This term describes an effective interaction leading to the dPI~\cite{Yamase_dPI_2, Khavkine_dPI, Kee_dPI}. 
Thus, the d-wave form factor $d_{\bm{k}} = \cos k_{x} - \cos k_{y}$ is adopted. 
In order to study the bilayer system, we take into account not only the intralayer forward scattering term 
but also the interlayer one with $g_{1}$ and $g_{2}$ representing the coupling constants. 
The EQ or EO state may be stabilized by the forward scattering term. 
 
%%%%%%%%%%%%%%%%%%%%%%%%%%%%%%%%%%%%%%%%%%%%%%%%%%%%%%%%%%%%%%%%%%%
%%%%%%%%%%%%%%%%%%%%%%%%%%%%%%%%%%%%%%%%%%%%%%%%%%%%%%%%%%%%%%%%%%%
%%%%%%%%%%%%%%%%%%%%%%%%%%%%%%%%%%%%%%%%%%%%%%%%%%%%%%%%%%%%%%%%%%%

%%%%%%%%%%%%%%%%%%%%%%%%%%%%%%%%%%%%%%%%%%%%%%%%%%%%%%%%%%%%%%%%%%%
%%%%%%%%%%%%%%%%%%%%%%%%%%%%%%%%%%%%%%%%%%%%%%%%%%%%%%%%%%%%%%%%%%%
%%%%%%%%%%%%%%%%%%%%%%%%%%%%%%%%%%%%%%%%%%%%%%%%%%%%%%%%%%%%%%%%%%%
\subsection{Mean field theory}
We apply the mean-field approximation to the forward scattering term $H_{f}$ in Eq. (\ref{eq:5}). 
By decoupling $n_{\bm{k} l} n_{\bm{k}' l'} \simeq  n_{\bm{k} l} \langle n_{\bm{k}' l'} \rangle + \langle n_{\bm{k}l} \rangle n_{\bm{k}' l'} - \langle n_{\bm{k} l} \rangle \langle n_{\bm{k}' l'} \rangle $, the order parameter of dPI on the $l$-layer is obtained as
\begin{equation}
\Delta_{l} = \Delta_{1 l} + \Delta_{2 \bar{l}} ,   \label{eq:6}
\end{equation}
where 
\begin{align}
\Delta_{1 l} &= - \frac{g_{1}}{N} \sum_{\bm{k}} d_{\bm{k}} \langle n_{\bm{k}l} \rangle,    \label{eq:7} \\
\Delta_{2 l} &= - \frac{g_{2}}{N} \sum_{\bm{k}} d_{\bm{k}} \langle n_{\bm{k}l} \rangle,    \label{eq:8}
\end{align} 
and $\bar{l}$ indicates the layer different from $l$, i.e., $\{l,\bar{l} \, \} = \{A, B\}$.
The intralayer and interlayer contributions to the order parameter $\Delta_{l}$ are represented  
by $\Delta_{1 l}$ and $\Delta_{2 \bar{l}}$, respectively. 
When the dPI order parameter $\Delta_{l}$ is finite, the Fermi surface deforms depending on the sign 
of $\Delta_l$ [see Fig.~\ref{fig:FSs_each_multipole_state}]. 
The EQ state is characterized by $(\Delta_{A}, \Delta_{B})=(\Delta,\Delta)$, while the EO state is characterized by 
$(\Delta_{A}, \Delta_{B})=(\Delta,-\Delta)$.

The model is reduced to the mean field Hamiltonian 
\begin{align}
H^{\rm{MF}} &= \sum_{\bm{k}} \hat{C}^{\dagger}_{\bm{k}} \ \hat{H}^{\, \rm{MF}}_{4} (\bm{k}) \ \hat{C}_{\bm{k}} + E_{\rm{cond}} , \label{eq:9} 
\end{align}
where 
\begin{align}
E_{\rm{cond}} &= \frac{N}{2g_{1}} \left[(\Delta_{1A})^{2} + (\Delta_{1B})^{2} \right] 
+ \frac{N}{g_{2}} \Delta_{2A} \Delta_{2B} , \label{eq:10}
\end{align}
and $\hat{C}^{\dagger}_{\bm{k}} = (c^{\dagger}_{\bm{k} \uparrow A}, c^{\dagger}_{\bm{k} \downarrow A}, c^{\dagger}_{\bm{k} \uparrow B}, c^{\dagger}_{\bm{k} \downarrow B})$ is a vector operator. 
The  $4 \times 4$ matrix $\hat{H}^{\, \rm{MF}}_{4} (\bm{k})$ is obtained as 
\begin{equation}
\hat{H}^{\, \rm{MF}}_{4} (\bm{k}) =
\scalebox{1.3}{$\displaystyle
{\footnotesize
\begin{pmatrix}
\xi_{\bm{k} A}                & -\alpha \lambda_{\bm{k}}^{+}      & t_{\perp}                         & 0 \\
-\alpha \lambda_{\bm{k}}^{-}  & \xi_{\bm{k} A}                    & 0                               & t_{\perp} \\
t_{\perp}                    & 0                              & \xi_{\bm{k} B}                     &  \alpha \lambda_{\bm{k}}^{+} \\
0                          & t_{\perp}                        &  \alpha \lambda_{\bm{k}}^{-}       & \xi_{\bm{k} B}  
\end{pmatrix} ,
} $}
\label{eq:11}
\end{equation}
where $\lambda_{\bm{k}}^{\pm} = \sin k_{y} \pm i \sin k_{x}$ and $\xi_{\bm{k} l} = \varepsilon_{\bm{k}} + d_{\bm{k}} \Delta_{l}$.
Performing a unitary transformation, 
\begin{align}
c_{\bm{k}sl} &= \sum_{\nu=1}^{4} u_{\bm{k}sl}^{\nu} \gamma_{\bm{k}\nu}, \label{eq:13}
\end{align}
we obtain the band representation 
\begin{align}
H^{\rm{MF}} &= \sum_{\bm{k}} \sum_{\nu=1}^{4} E_{\bm{k} \nu} \gamma_{\bm{k}\nu}^{\dagger} \gamma_{\bm{k}\nu} + E_{\rm{cond}} ,  
\label{eq:12} 
\end{align}
with $E_{\bm{k} \nu}$ being a quasiparticle's energy.  
Equations~(\ref{eq:7}) and (\ref{eq:8}) are recast into 
\begin{align}
\Delta_{1 l} &= - \frac{g_{1}}{N} \sum_{\bm{k}, s}  \sum_{\nu = 1}^{4} d_{\bm{k}} |u_{\bm{k}sl}^{\nu}|^{2} f(E_{\bm{k} \nu}),    \label{eq:7'} \\
\Delta_{2 l} &= - \frac{g_{2}}{N} \sum_{\bm{k}, s}  \sum_{\nu = 1}^{4} d_{\bm{k}} |u_{\bm{k}sl}^{\nu}|^{2} f(E_{\bm{k} \nu}),    \label{eq:8'}
\end{align}
where $f(E)$ is the Fermi--Dirac distribution function. 
Equations~(\ref{eq:7'}) and (\ref{eq:8'}) are self-consistent equations to be solved numerically. 
The free energies of the (meta)stable normal state, EQ state, and EO state are calculated on the basis of the 
mean field Hamiltonian, Eq.~(\ref{eq:12}). 
%%%%%%%%%%%%%%%%%%%%%%%%%%%%%%%%%%%%%%%%%%%%%%%%%%%%%%%%%%%%%%%%%%%
%%%%%%%%%%%%%%%%%%%%%%%%%%%%%%%%%%%%%%%%%%%%%%%%%%%%%%%%%%%%%%%%%%%
%%%%%%%%%%%%%%%%%%%%%%%%%%%%%%%%%%%%%%%%%%%%%%%%%%%%%%%%%%%%%%%%%%%

%%%%%%%%%%%%%%%%%%%%%%%%%%%%%%%%%%%%%%%%%%%%%%%%%%%%%%%%%%%%%%%%%%%
%%%%%%%%%%%%%%%%%%%%%%%%%%%%%%%%%%%%%%%%%%%%%%%%%%%%%%%%%%%%%%%%%%%
%%%%%%%%%%%%%%%%%%%%%%%%%%%%%%%%%%%%%%%%%%%%%%%%%%%%%%%%%%%%%%%%%%%
\section{Multipole Susceptibility}

When the multipole order occurs through the second-order phase transition, the critical point 
is given by the divergence of multipole susceptibility. Therefore, it is useful to calculate the multipole susceptibility 
in order to examine the thermodynamical stability of the multipole states. 
In this section, an analytic form of the multipole susceptibility is obtained, and  
the effects of the ASOC are clarified.

\subsection{Random phase approximation}

Following previous works \cite{Adachi_multipole_susceptibility, Yamase_multipole_susceptibility, Dell'Anna_multipole_susceptibility}, we define the susceptibility in the spin- and layer-dependent form 
\begin{equation}
\chi^{\, d}_{sl, s'l'} (\bm{q}, i \omega_{n}) = \frac{1}{N} \int^{1/T}_{0} d \tau e^{i \omega_{n} \tau} \langle n^{\, d}_{sl} (\bm{q},\tau) n^{\, d}_{s'l'} (-\bm{q},0) \rangle , \label{eq:15}
\end{equation}
where $\omega_{n} = 2 n \pi T$ are boson Matsubara frequencies, and 
$n^{\, d}_{sl} (\bm{q}) = \sum_{\bm{k}} d_{\bm{k}} \, c^{\dagger}_{\bm{k} + \bm{q}/2 , s, l} \, c_{\bm{k} - \bm{q}/2 , s, l}$ is 
the d-wave density operator. 
We now consider the forward scattering process and calculate the uniform and static susceptibility 
in the limit $(\bm{q}, \omega_{n}) \rightarrow (\bm{0}, 0)$. 
The irreducible susceptibility is obtained as 
\begin{align}
\chi^{\, d,0}_{sl, s'l'} (\bm{0}, 0) &= -\frac{1}{N} \sum_{\bm{k},\nu,\nu'} \lim_{\bm{q} \to \bm{0}} d_{\bm{k}} d_{\bm{k}+\bm{q}} \notag \\ 
 & \hspace{12mm} \times \frac{f(E_{\bm{k} \nu'}) - f(E_{\bm{k}+\bm{q} \nu})}{E_{\bm{k} \nu'} - E_{\bm{k}+\bm{q} \nu}} A^{\nu \nu'}_{\bm{k},sl,s'l'} (\bm{q}), \label{eq:16}
\end{align}
and the coefficients $A^{\nu \nu'}_{\bm{k},sl,s'l'} (\bm{q})$ are given by 
\begin{equation}
A^{\nu \nu'}_{\bm{k},sl,s'l'} (\bm{q})  =  u^{\nu}_{\bm{k}+\bm{q} s l}  \, u^{\nu *}_{\bm{k}+\bm{q} s' l'} \, u^{\nu'}_{\bm{k} s' l'} \, u^{\nu' *}_{\bm{k} s l}, \label{eq:17}
\end{equation}
where $E_{\bm{k} \nu}$ and $u_{\bm{k}sl}^{\nu}$ are defined for the noninteracting Hamiltonian  
$H_{0} = H_{\rm{kin}} + H_{\rm{ASOC}} + H_{\perp}$.

For the electric multipole order, it is sufficient to calculate the spin-independent part 
\begin{equation}
\chi^{\, d}_{l l'} (\bm{q}, i \omega_{n})  =  \sum_{s,s'} \chi^{\, d}_{sl, s'l'} (\bm{q}, i \omega_{n}),  \label{eq:14}
\end{equation}
which is nothing but the susceptibility of the electric multipole. 
By using the random phase approximation, we obtain the multipole susceptibility in matrix form 
\begin{eqnarray}
\hat{\chi}^{\, d} (\bm{0}, 0) &=&
\scalebox{1.2}{$\displaystyle
{\footnotesize
\begin{pmatrix}
\chi^{\, d}_{A A} (\bm{0}, 0)  & \chi^{\, d}_{A B} (\bm{0}, 0) \\[7pt]
\chi^{\, d}_{B A} (\bm{0}, 0)  & \chi^{\, d}_{B B} (\bm{0}, 0)
\end{pmatrix} 
}$} \label{eq:22} 
\\
&=& \frac{\hat{\chi}^{\, d,0} (\bm{0}, 0)}{\hat{1} + \hat{g} \ \hat{\chi}^{\, d,0} (\bm{0}, 0)}. \label{eq:23}
\end{eqnarray}
The matrix element of the irreducible susceptibility, $\hat{\chi}^{\, d,0} (\bm{0}, 0)$, is obtained by 
taking the summation for spin indices, 
\begin{equation}
\chi^{\, d,0}_{l l'} (\bm{0}, 0) = \sum_{s,s'} \chi^{\, d,0}_{sl, s'l'} (\bm{0}, 0). \label{eq:21}
\end{equation}
The $2 \times 2$ matrix $\hat{g}$ is given by
\begin{equation}
\hat{g} =
\scalebox{1.2}{$\displaystyle
{\footnotesize
\begin{pmatrix}
-g_{1}  & -g_{2} \\[5pt]
-g_{2}  & -g_{1}
\end{pmatrix}. 
}$} \label{eq:24}    
\end{equation}

%\begin{equation}
%\hat{\chi}^{d} (\bm{0}, 0) =
%\scalebox{1.2}{$\displaystyle
%{\footnotesize
%\begin{pmatrix}
%\chi^{d}_{A A} (\bm{0}, 0)  & \chi^{d}_{A B} (\bm{0}, 0) \\[7pt]
%\chi^{d}_{B A} (\bm{0}, 0)  & \chi^{d}_{B B} (\bm{0}, 0)
%\end{pmatrix}.
%}$} \label{eq:22}
%\end{equation}

The second-order multipole order occurs when the multipole susceptibility diverges. 
Thus, the eigenvalue of $- \hat{g} \, \hat{\chi}^{\, d,0} (\bm{0}, 0)$ is unity at the critical point. 
Diagonalizing $- \hat{g} \, \hat{\chi}^{\, d,0} (\bm{0}, 0)$, we obtain two eigenvalues; 
\begin{align}
\lambda_{\rm{EO}} = (g_{1} - g_{2}) \hspace{1mm} \Bigl(\chi^{\, d,0}_{AA}  - \chi^{\, d,0}_{AB} \Bigr) \hspace{8mm} \rm{for} \hspace{2mm} \rm{EO} \hspace{1mm} \rm{order},   \label{eq:25} \\
\lambda_{\rm{EQ}} = (g_{1} + g_{2}) \hspace{1mm} \Bigl(\chi^{\, d,0}_{AA}  + \chi^{\, d,0}_{AB} \Bigr) \hspace{8mm} \rm{for} \hspace{2mm} \rm{EQ} \hspace{1mm} \rm{order}.   \label{eq:26}
\end{align}
We used the relations $\chi^{\, d,0}_{AA} = \chi^{\, d,0}_{BB}$ and $\chi^{\, d,0}_{AB} = \chi^{\, d,0}_{BA}$.
The EO (EQ) order occurs when $\lambda_{\rm{EO}} = 1$ ($\lambda_{\rm{EQ}} = 1$).
%%%%%%%%%%%%%%%%%%%%%%%%%%%%%%%%%%%%%%%%%%%%%%%%%%%%%%%%%%%%%%%%%%%
%%%%%%%%%%%%%%%%%%%%%%%%%%%%%%%%%%%%%%%%%%%%%%%%%%%%%%%%%%%%%%%%%%%
%%%%%%%%%%%%%%%%%%%%%%%%%%%%%%%%%%%%%%%%%%%%%%%%%%%%%%%%%%%%%%%%%%%

%%%%%%%%%%%%%%%%%%%%%%%%%%%%%%%%%%%%%%%%%%%%%%%%%%%%%%%%%%%%%%%%%%%
%%%%%%%%%%%%%%%%%%%%%%%%%%%%%%%%%%%%%%%%%%%%%%%%%%%%%%%%%%%%%%%%%%%
%%%%%%%%%%%%%%%%%%%%%%%%%%%%%%%%%%%%%%%%%%%%%%%%%%%%%%%%%%%%%%%%%%%
\subsection{Analytic form of irreducible susceptibility}
In this subsection, we show the analytic results of the irreducible susceptibilities $\chi^{\, d,0}_{A A} (\bm{0}, 0)$ 
and $\chi^{\, d,0}_{A B} (\bm{0}, 0)$ by calculating Eqs.~(\ref{eq:16}), (\ref{eq:17}), and (\ref{eq:21}). 
As a result of the calculation in Appendix A, we obtain 
\begin{align}
\chi&^{\, d,0}_{AA}  (\bm{0} , 0) = \frac{1}{N} \sum_{\bm{k}}  d^{\, 2}_{\bm{k}} \notag \\
& \times \Biggl[\frac{1}{4T}  \biggl \{ T^{4}_{\bm{k}} + (1-T^{2}_{\bm{k}})^{2} \biggr \} \biggl \{ \frac{1}{\cosh^{2} (E_{\bm{k}1}/2T)} + \frac{1}{\cosh^{2} (E_{\bm{k}3}/2T)} \biggr \} \notag \\
  & \hspace{12mm} + \frac{ T^{2}_{\bm{k}} (1 - T^{2}_{\bm{k}}) }{ \sqrt{(\alpha |\bm{g}_{\bm{k}}|)^{2} + t^{2}_{\perp}} } \biggl \{\tanh (E_{\bm{k}1}/2T) - \tanh (E_{\bm{k}3}/2T) \biggr \} \Biggr] , \label{eq:27}
\end{align}
and 
\begin{align}
\chi^{\, d,0}_{AB}  & (\bm{0} , 0) =  \frac{1}{N}  \sum_{\bm{k}}  d^{\, 2}_{\bm{k}} T^{2}_{\bm{k}} (1 - T^{2}_{\bm{k}}) \notag \\
& \times \Biggl[\frac{1}{2T}  \biggl \{ \frac{1}{\cosh^{2} (E_{\bm{k}1}/2T)} + \frac{1}{\cosh^{2} (E_{\bm{k}3}/2T)} \biggr \} \notag \\
  & \hspace{7mm} - \frac{1}{ \sqrt{(\alpha |\bm{g}_{\bm{k}}|)^{2} + t^{2}_{\perp}} } \biggl \{\tanh (E_{\bm{k}1}/2T) - \tanh (E_{\bm{k}3}/2T) \biggr \} \Biggr] , \label{eq:28}
\end{align}
where $|\bm{g}_{\bm{k}}| = ( \sin^{2} k_{x} + \sin^{2} k_{y} )^{1/2}$ is the magnitude of the Rashba g-vector. 
The dispersion relation in the $\nu$-th eigenstate of the noninteracting Hamiltonian $H_{0}$ is represented as
\begin{align}
E_{\bm{k} 1} &= E_{\bm{k} 2} = \varepsilon_{\bm{k}} + \sqrt{(\alpha |\bm{g}_{\bm{k}}|)^{2} + t^{2}_{\perp}} , \label{eq:29} \\
E_{\bm{k} 3} &= E_{\bm{k} 4} = \varepsilon_{\bm{k}} - \sqrt{(\alpha |\bm{g}_{\bm{k}}|)^{2} + t^{2}_{\perp}} . \label{eq:30}
\end{align}
We introduced $T_{\bm{k}}$ given by 
\begin{equation}
T_{\bm{k}} \equiv \frac{t_{\perp}}{\sqrt{t^{2}_{\perp} + \left[\alpha |\bm{g}_{\bm{k}}| + \sqrt{ (\alpha |\bm{g}_{\bm{k}}|)^{2} + t^{2}_{\perp}} \ \right]^{2}}}. \label{eq:31}
\end{equation}
Notice that the first and second terms in Eqs. (\ref{eq:27}) and (\ref{eq:28}) come from 
the intraband and interband contributions, respectively.

\subsection{Effect of spin-orbit coupling}

Now the effects of the Rashba ASOC on the irreducible susceptibility are elucidated.
For this purpose, we discuss the two limiting cases $\alpha/t_{\perp} =0$ and $\alpha/t_{\perp} = \infty$. 

In the absence of the ASOC, $\alpha/t_{\perp} =0$, Eq.~(\ref{eq:31}) is reduced to $T_{\bm{k}} = 1/\sqrt{2}$.
Then, the irreducible susceptibilities are represented as 
\begin{align}
\chi^{\, d,0}_{AA}  (\bm{0} , 0) = & \frac{1}{4N}  \sum_{\bm{k}}  d^{\, 2}_{\bm{k}} \notag \\
& \times \Biggl[\frac{1}{2T} \biggl \{ \frac{1}{\cosh^{2} (E_{\bm{k}1}/2T)} + \frac{1}{\cosh^{2} (E_{\bm{k}3}/2T)} \biggr \} \notag \\
  & \hspace{5mm} + \frac{ 1 }{ t_{\perp} }  \biggl \{\tanh (E_{\bm{k}1}/2T) - \tanh (E_{\bm{k}3}/2T) \biggr \} \Biggr] , \label{eq:32}
\end{align}
\begin{align}
\chi^{\, d,0}_{AB}  (\bm{0} , 0) = & \frac{1}{4N}  \sum_{\bm{k}}  d^{\, 2}_{\bm{k}} \notag \\
& \times \Biggl[\frac{1}{2T} \biggl \{ \frac{1}{\cosh^{2} (E_{\bm{k}1}/2T)} + \frac{1}{\cosh^{2} (E_{\bm{k}3}/2T)} \biggr \} \notag \\
  & \hspace{5mm} - \frac{ 1 }{ t_{\perp} }  \biggl \{\tanh (E_{\bm{k}1}/2T) - \tanh (E_{\bm{k}3}/2T) \biggr \} \Biggr] , \label{eq:33}
\end{align}
where $E_{\bm{k} 1} = \varepsilon_{\bm{k}} + t_{\perp}$ and $E_{\bm{k} 3} = \varepsilon_{\bm{k}} - t_{\perp}$. 
We see that the intraband contributions to $\chi^{\, d,0}_{AA}$ and $\chi^{\, d,0}_{AB}$ 
[first term in Eqs.~(\ref{eq:32}) and (\ref{eq:33})] are equivalent, while the interband contributions 
[second term in Eqs.~(\ref{eq:32}) and (\ref{eq:33})] are opposite. 
Although the intralayer irreducible susceptibility $\chi^{\, d,0}_{AA}$ is always positive, 
the interlayer one $\chi^{\, d,0}_{AB}$ may be negative owing to the interband contribution. 
Because at low temperatures the intraband contribution is proportional to the DOS at the Fermi level, 
$\chi^{\, d,0}_{AB}$ is positive when the Fermi level is in the vicinity of the van Hove singularity. 
Then, the EQ state is favored according to Eqs.~(\ref{eq:25}) and (\ref{eq:26}). 
This is the situation that was studied previously\cite{Yamase_Bilayer_1,Yamase_Bilayer_2}.

In the opposite limit $\alpha/t_{\perp} = \infty$, the Rashba ASOC is much larger than the interlayer hopping amplitude, 
and $T_{\bm{k}} \rightarrow 0$. 
Then, we obtain the simple form  
\begin{align}
\chi^{\, d,0}_{AA}  (\bm{0} , 0) &= \frac{1}{4NT} \sum_{\bm{k}}  d^{\, 2}_{\bm{k}} \biggl \{ \frac{1}{\cosh^{2} (E_{\bm{k}1}/2T)} + \frac{1}{\cosh^{2} (E_{\bm{k}3}/2T)} \biggr \} , \label{eq:34} \\
\chi^{\, d,0}_{AB}  (\bm{0} , 0) &= 0, \label{eq:35}
\end{align}
where $E_{\bm{k} 1} = \varepsilon_{\bm{k}} + \alpha |\bm{g}_{\bm{k}}|$ and $E_{\bm{k} 3} = \varepsilon_{\bm{k}} - \alpha |\bm{g}_{\bm{k}}|$.
Interestingly, the interlayer irreducible susceptibility $\chi^{\, d,0}_{AB}(\bm{0}, 0)$ vanishes in the large ASOC limit. 
This result is reasonable because the interlayer irreducible susceptibility comes from the interlayer kinetic energy 
that is suppressed by the ASOC~\cite{Maruyama_multilayer_SC}.
Thus, the kinetic contribution to the interlayer coupling between the dPI order parameters $\Delta_A$ and $\Delta_B$ 
disappears when $\alpha/t_{\perp} \gg 1$. 
Then, the relative stability of the EO and EQ states is determined by the forward scattering interaction. 
Indeed, the magnitude relation between the two eigenvalues $\lambda_{\rm{EO}}$ and $\lambda_{\rm{EQ}}$ is determined 
by the sign of the interlayer forward scattering interaction $g_2$. 
Notice that the main ingredients stabilizing the EO or EQ state are different 
between the small and large ASOC regions.

%%%%%%%%%%%%%%%%%%%%%%%%%%%%%%%%%%%%%%%%%%%%%%%%%%%%%%%%%%%%%%%%%%%
%%%%%%%%%%%%%%%%%%%%%%%%%%%%%%%%%%%%%%%%%%%%%%%%%%%%%%%%%%%%%%%%%%%
%%%%%%%%%%%%%%%%%%%%%%%%%%%%%%%%%%%%%%%%%%%%%%%%%%%%%%%%%%%%%%%%%%%
%%%%%%%%%%%%%%%%%%%%%%%%%%%%%%%%%%%%%%%%%%%%%%%%%%%%%%%%%%%%%%%%%%%
%%%%%%%%%%%%%%%%%%%%%%%%%%%%%%%%%%%%%%%%%%%%%%%%%%%%%%%%%%%%%%%%%%%

%%%%%%%%%%%%%%%%%%%%%%%%%%%%%%%%%%%%%%%%%%%%%%%%%%%%%%%%%%%%%%%%%%%
%%%%%%%%%%%%%%%%%%%%%%%%%%%%%%%%%%%%%%%%%%%%%%%%%%%%%%%%%%%%%%%%%%%
%%%%%%%%%%%%%%%%%%%%%%%%%%%%%%%%%%%%%%%%%%%%%%%%%%%%%%%%%%%%%%%%%%%
%%%%%%%%%%%%%%%%%%%%%%%%%%%%%%%%%%%%%%%%%%%%%%%%%%%%%%%%%%%%%%%%%%%
%%%%%%%%%%%%%%%%%%%%%%%%%%%%%%%%%%%%%%%%%%%%%%%%%%%%%%%%%%%%%%%%%%%

\section{Numerical Results}
%%%%%%%%%%%%%%%%%%%%%%%%%%%%%%%%%%%%%%%%%%%%%%%%%%%%%%%%%%%%%%%%%%%
%%%%%%%%%%%%%%%%%%%%%%%%%%%%%%%%%%%%%%%%%%%%%%%%%%%%%%%%%%%%%%%%%%%
%%%%%%%%%%%%%%%%%%%%%%%%%%%%%%%%%%%%%%%%%%%%%%%%%%%%%%%%%%%%%%%%%%%
\subsection{Phase diagram}
In this section, we show numerical results of the mean field theory. 
First, the phase diagram of multipole states is discussed. 
Our calculations reproduce the previous results in the absence of the ASOC and interlayer interaction\cite{Yamase_Bilayer_1,Yamase_Bilayer_2}. At $\alpha=g_2=0$, the EQ state is stabilized in the weak-coupling regime, for instance, at $g_{1} = 0.5$. 
We examine the effect of the layer-dependent Rashba ASOC below.

\begin{figure}[htbp]
 \begin{center}
\includegraphics[width=0.5\hsize]{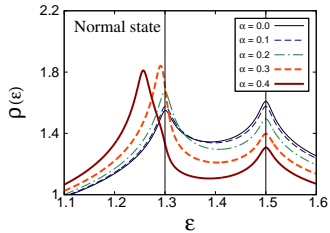}
\caption{(Color online) DOS $\rho(\varepsilon)$ for several values of $\alpha$ in the normal state, 
i.e., $\Delta_{A} = \Delta_{B} = 0$. We assume $\mu=0$. 
}
\label{fig:DOS}
  \end{center}
\end{figure}

For discussions of the phase diagram, we calculate the density of states (DOS),  
$\rho(\varepsilon) = \frac{1}{N} \sum_{\bm{k},\nu} \delta (\varepsilon - E_{\bm{k} \nu})$, in the normal state. 
Figure~\ref{fig:DOS} shows the DOS for various magnitudes of the ASOC. 
The DOS at the Fermi energy $\rho(0)$ is large for the chemical potential $\mu \simeq 1.3$ or $1.5$ 
because of the van Hove singularity at ${\bm k} = (\pm \pi,0)$ and $(0, \pm \pi)$. 
The Fermi surface of the higher band with $E_{\bm{k} 1} = E_{\bm{k} 2}$ (lower band with $E_{\bm{k} 3} = E_{\bm{k} 4}$) 
crosses the van Hove singularity when $\mu \simeq 1.5$ ($\mu \simeq 1.3$). 
With increasing the ASOC, DOS at the peak at approximately $\mu \simeq 1.3$ is enhanced, while the DOS 
at around $\mu \simeq 1.5$ is suppressed. 
We also see the shift of the peak at around $\mu \simeq 1.3$ to the low-energy region.

\begin{figure}[htbp]
 \begin{center}
\includegraphics[width=0.6\hsize]{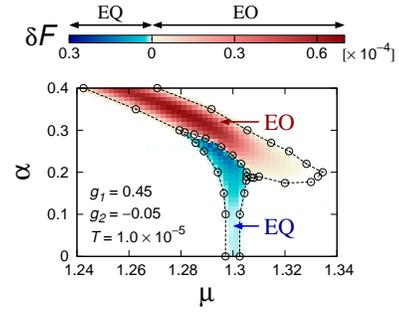}
\caption{(Color online) Phase diagram of multipole states in the $\mu$-$\alpha$ plane at $T = 1.0 \times 10^{-5}$.
                        We choose the intralayer and interlayer forward scatterings,  
                        $g_{1} = 0.45$ and $g_{2} = -0.05$, respectively.
                        The red (blue) region shows the EO (EQ) phase. 
                        The contrasting density indicates the condensation energy $\delta F$ of EO and EQ states. 
                        The dashed lines with circles show the first-order phase boundary. 
}
\label{fig:FIG_cp_ASOC}
  \end{center}
\end{figure}

The dPI is likely to occur when the DOS is large~\cite{Yamase_dPI_1, Yamase_dPI_2, Khavkine_dPI, Kee_dPI,Yamase_Bilayer_1,Yamase_Bilayer_2}. 
Therefore, it is expected that the multipole order is enhanced at around $\mu \simeq 1.3$ by the ASOC. 
Thus, we investigate the phase diagram at around $\mu \simeq 1.3$ and show Fig.~\ref{fig:FIG_cp_ASOC}. 
The condensation energy of multipole states defined by $\delta F = F(0, 0) - F(\Delta_A, \Delta_B)$  
with $F(\Delta_A, \Delta_B)$ being the free energy is plotted. 
As we have expected, the multipole order is enhanced by the ASOC. 
It is also shown that the multipole phase shifts to the low-energy region following the peak of the DOS.  
On the other hand, we confirmed that the multipole order is suppressed at around  $\mu \simeq 1.5$ by switching on the ASOC.

In the small ASOC region, the EQ state rather than the EO state gains kinetic energy, as indicated 
by the positive $\chi^{\, d,0}_{AB}(\bm{0}, 0)$. 
As we have shown in Sect.~4.3, this property is universal when the Fermi surface is close to the van Hove singularity. 
Actually, the EQ state is stable in the small ASOC region of Fig.~\ref{fig:FIG_cp_ASOC}.

On the other hand, the odd-parity EO state is stabilized by a moderate ASOC. 
This is because $\chi^{\, d,0}_{AB}(\bm{0}, 0)$ is suppressed, and a weak interlayer forward scattering $g_2 <0$ 
stabilizes the EO state rather than the EQ state. 
The sign of $g_2$ depends on the microscopic details of the system and is beyond the scope of this paper. 
However, we can say that the layer-dependent Rashba ASOC favors the odd-parity EO order 
by suppressing the kinetic energy gain of the EQ state. 
This is one of the main findings of this paper.

%%%%%%%%%%%%%%%%%%%%%%%%%%%%%%%%%%%%%%%%%%%%%%%%%%%%%%%%%%%%%%%%%%%
%%%%%%%%%%%%%%%%%%%%%%%%%%%%%%%%%%%%%%%%%%%%%%%%%%%%%%%%%%%%%%%%%%%
%%%%%%%%%%%%%%%%%%%%%%%%%%%%%%%%%%%%%%%%%%%%%%%%%%%%%%%%%%%%%%%%%%%

\begin{figure}[htbp]
  \begin{center}
\vspace{8mm}
    \includegraphics[width=0.50\hsize]{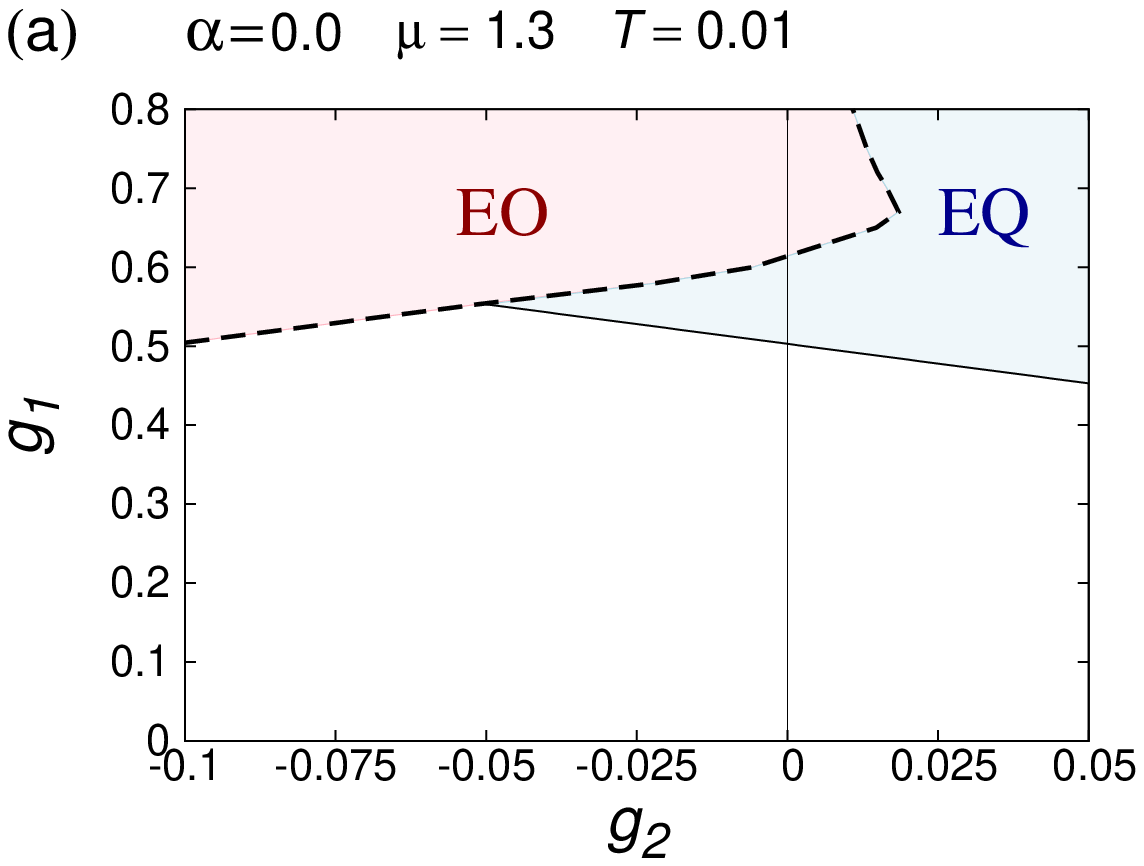}
\\ \vspace{8mm}
    \includegraphics[width=0.50\hsize]{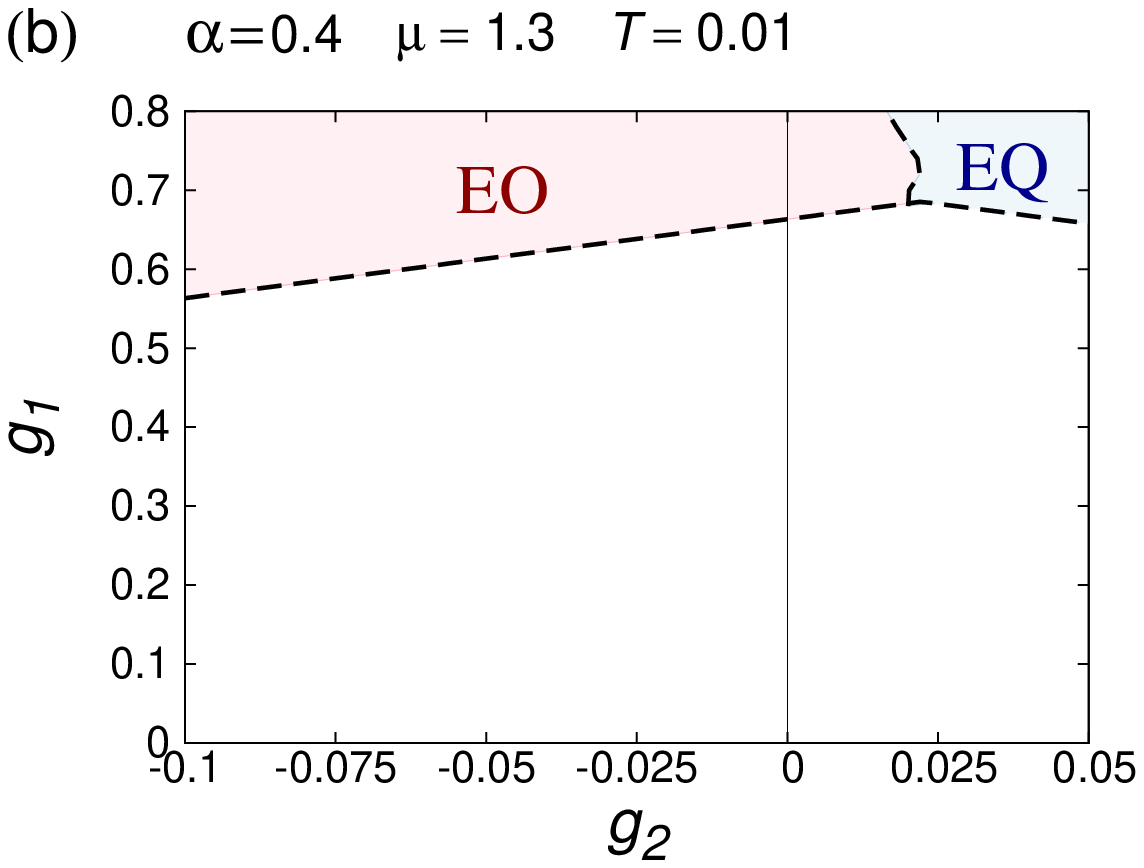}
    \caption{(Color online) Phase diagram in the $g_1$-$g_2$-plane at $T=0.01$ and $\mu=1.3$. 
    (a) $\alpha=0$ and (b) $\alpha=0.4$. The first-order transition and second-order transition are shown by 
    the dashed line and solid line, respectively. 
    }
    \label{fig:FIG_g1_g2_PHASE_DIAGRAM}
 \end{center}
\end{figure}

Figure~\ref{fig:FIG_g1_g2_PHASE_DIAGRAM} shows the phase diagram in the $g_1$-$g_2$-plane at $T=0.01$. 
In the absence of the ASOC, the phase boundary between the EO and EQ states significantly depends on the 
magnitude of $g_1$ [Fig.~\ref{fig:FIG_g1_g2_PHASE_DIAGRAM}(a)]. 
The EQ state is stable in the weak coupling region, while the EO state may be stable 
in the strong coupling region~\cite{Yamase_Bilayer_1,Yamase_Bilayer_2}. 
On the other hand, the stability of the EQ and EO states is almost independent of $g_1$ and 
determined by $g_2$ in the large ASOC region, as shown by Fig.~\ref{fig:FIG_g1_g2_PHASE_DIAGRAM}(b). 
Thus, the discussions in this subsection are confirmed. 
For the parameters in Fig.~\ref{fig:FIG_g1_g2_PHASE_DIAGRAM}(b), the EO state is stable even in the absence of 
the interlayer interaction, namely, $g_2=0$.

%%%%%%%%%%%%%%%%%%%%%%%%%%%%%%%%%%%%%%%%%%%%%%%%%%%%%%%%%%%%%%%%%%%
%%%%%%%%%%%%%%%%%%%%%%%%%%%%%%%%%%%%%%%%%%%%%%%%%%%%%%%%%%%%%%%%%%%
%%%%%%%%%%%%%%%%%%%%%%%%%%%%%%%%%%%%%%%%%%%%%%%%%%%%%%%%%%%%%%%%%%%
\subsection{Multipole susceptibility}
\begin{figure}[htbp]
 \begin{center}
   \includegraphics[width=0.5\hsize]{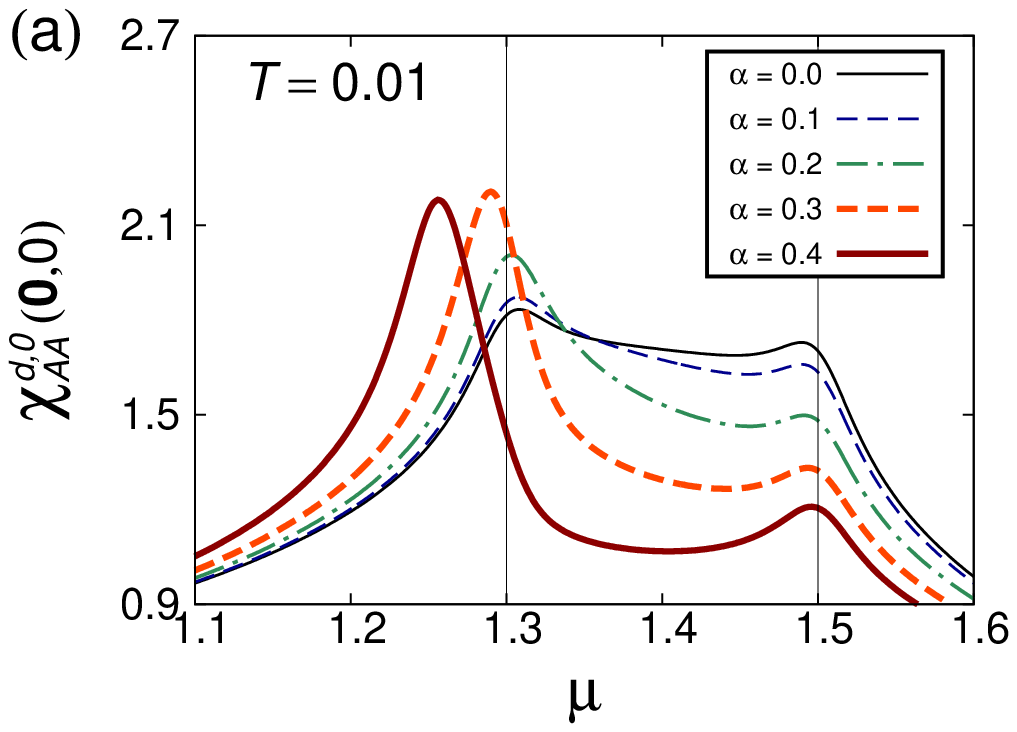}
   \includegraphics[width=0.5\hsize]{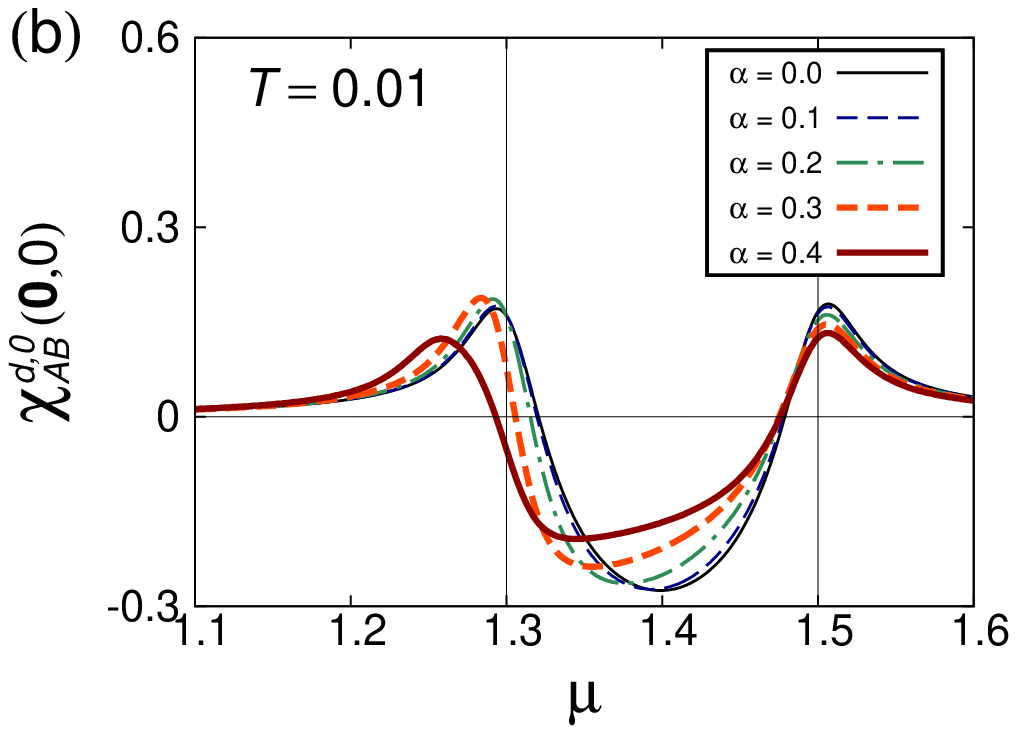}
   \caption{(Color online) Irreducible susceptibilities (a) $\chi^{\, d,0}_{A A} (\bm{0}, 0)$ and (b) $\chi^{\, d,0}_{A B} (\bm{0}, 0)$ 
at $T=0.01$ as a function of $\mu$ for several values of $\alpha$.
We confirmed $\chi^{\, d,0}_{A A} = \chi^{\, d,0}_{B B}$ and $\chi^{\, d,0}_{A B} = \chi^{\, d,0}_{B A}$. 
   }
   \label{fig:chi_T_0.01}
 \end{center}
\end{figure}

In the previous subsection, we discussed the $\mu$-$\alpha$ phase diagram on the basis of the analytic form 
of the multipole susceptibility obtained in Sect.~4. 
Here we numerically estimate the multipole susceptibility.
Figure \ref{fig:chi_T_0.01} shows the chemical potential dependence of the irreducible multipole susceptibilities $\chi^{\, d,0}_{A A}$ and $\chi^{\, d,0}_{A B}$.

As shown in Fig. \ref{fig:chi_T_0.01}(a), the intralayer susceptibility $\chi^{\, d,0}_{A A}$ shows peaks at $\mu$ 
on the van Hove singularity. In the large ASOC region, the $\mu$ dependence of $\chi^{\, d,0}_{A A}$ resembles 
that of DOS in agreement with Eq.~(\ref{eq:34}). It is confirmed that the multipole order 
is enhanced by the ASOC at around $\mu \simeq 1.3$, although it is suppressed at around $\mu \simeq 1.5$. 
In the absence of the ASOC, a nearly flat $\mu$ dependence of $\chi^{\, d,0}_{A A}$ appears between $\mu=1.3$ and $1.5$ 
because of the interband contribution [second term in Eq.~(\ref{eq:32})].

Figure \ref{fig:chi_T_0.01}(b) shows the sign change of the interlayer irreducible susceptibility $\chi^{\, d,0}_{A B}$ 
as a function of $\mu$. Although  $\chi^{\, d,0}_{A B}$ is positive around the van Hove singularities, 
it becomes negative at around $\mu = 1.4$, indicating the negative correlation of $\Delta_{A}$ and $\Delta_B$. 
This is consistent with the previous studies \cite{Yamase_Bilayer_1,Yamase_Bilayer_2} 
that showed the EO state at those chemical potentials. 
With increasing $\alpha$, the magnitude of $\chi^{\, d,0}_{A B}$ decreases as expected from Eq.~(\ref{eq:35}). 
In the large ASOC region, the single particle wave function is almost localized on 
each layer~\cite{Maruyama_multilayer_SC}, and thus, the kinetic energy is almost independent of 
the interlayer stacking of dPI order parameters. 
Therefore, the relative stability of the EQ and EO states is determined by the interlayer interaction energy $g_2$, 
as we discussed in Sect.~5.1.

%%%%%%%%%%%%%%%%%%%%%%%%%%%%%%%%%%%%%%%%%%%%%%%%%%%%%%%%%%%%%%%%%%%
%%%%%%%%%%%%%%%%%%%%%%%%%%%%%%%%%%%%%%%%%%%%%%%%%%%%%%%%%%%%%%%%%%%
%%%%%%%%%%%%%%%%%%%%%%%%%%%%%%%%%%%%%%%%%%%%%%%%%%%%%%%%%%%%%%%%%%%

%%%%%%%%%%%%%%%%%%%%%%%%%%%%%%%%%%%%%%%%%%%%%%%%%%%%%%%%%%%%%%%%%%%
%%%%%%%%%%%%%%%%%%%%%%%%%%%%%%%%%%%%%%%%%%%%%%%%%%%%%%%%%%%%%%%%%%%
%%%%%%%%%%%%%%%%%%%%%%%%%%%%%%%%%%%%%%%%%%%%%%%%%%%%%%%%%%%%%%%%%%%
\subsection{Second-order multipole transition} 
The eigenvalues $\lambda_{\rm EO}$ and $\lambda_{\rm EQ}$ defined by Eqs. (\ref{eq:25}) and (\ref{eq:26}) indicate 
the second-order multipole transition. 
For clarity, we show the analytic form obtained by Eqs.~(\ref{eq:27}) and (\ref{eq:28}), 
\begin{align}
\lambda&_{\rm{EO}} = \frac{g_{1} - g_{2}}{N} \sum_{\bm{k}}  d^{2}_{\bm{k}} \notag \\
& \times \Biggl[\frac{1}{4T}  ( 2 T^{2}_{\bm{k}} - 1 )^{2} \biggl \{ \frac{1}{\cosh^{2} (E_{\bm{k}1}/2T)} + \frac{1}{\cosh^{2} (E_{\bm{k}3}/2T)} \biggr \} \notag \\
       & \hspace{0mm} + \frac{ 2 T^{2}_{\bm{k}} (1 - T^{2}_{\bm{k}}) }{ \sqrt{(\alpha |\bm{g}_{\bm{k}}|)^{2} + t^{2}_{\perp}} } \biggl \{\tanh (E_{\bm{k}1}/2T) - \tanh (E_{\bm{k}3}/2T) \biggr \} \Biggr] , \label{eq:36} \\
\lambda&_{\rm{EQ}} = \frac{g_{1} + g_{2}}{4NT} \sum_{\bm{k}}  d^{2}_{\bm{k}} \biggl \{ \frac{1}{\cosh^{2} (E_{\bm{k}1}/2T)} + \frac{1}{\cosh^{2} (E_{\bm{k}3}/2T)} \biggr \}. \label{eq:37}
\end{align}
Equation~(\ref{eq:37}) reveals that the eigenvalue for the EQ state, $\lambda_{\rm EQ}$, is affected by the ASOC 
only through the energy spectrum. 

In the absence of the ASOC, $\alpha = 0$, $\lambda_{\rm EO}$ is reduced to 
\begin{align}
\lambda_{\rm{EO}} &= \frac{g_{1} - g_{2}}{2 N t_{\perp}}  \sum_{\bm{k}} d^{2}_{\bm{k}} \biggl \{\tanh (E_{\bm{k}1}/2T) - \tanh (E_{\bm{k}3}/2T) \biggr \} , \label{eq:38} 
%\\
%\lambda_{\rm{EQ}} &= \frac{g_{1} + g_{2}}{4 N T}  \sum_{\bm{k}} d^{2}_{\bm{k}} \biggl \{ \frac{1}{\cosh^{2} (E_{\bm{k}1}/2T)} + \frac{1}{\cosh^{2} (E_{\bm{k}3}/2T)} \biggr \}. \label{eq:39}
\end{align}
indicating that the EO order is triggered by the interband contribution, although the EQ order is caused 
by the intraband contribution. 
On the other hand, in the opposite limit $\alpha/t_{\perp} = \infty$, $\lambda_{\rm{EO}}$ is represented 
by the same form as $\lambda_{\rm{EQ}}$ except for the coupling constant $g_1 \pm g_2$, 
\begin{align}
\lambda_{\rm{EO}} & = \frac{g_{1} - g_{2}}{4NT} \sum_{\bm{k}}  d^{2}_{\bm{k}} \biggl \{ \frac{1}{\cosh^{2} (E_{\bm{k}1}/2T)} + \frac{1}{\cosh^{2} (E_{\bm{k}3}/2T)} \biggr \}.  
\label{eq:40} 
%\notag \\
%                & =(g_{1} - g_{2})  \hspace{1mm} \chi^{\, d,0}_{AA}  (\bm{0} , 0), \label{eq:40} 
%\\
%                & \notag \\
%\lambda_{\rm{EQ}} &=(g_{1} + g_{2})  \hspace{1mm} \chi^{\, d,0}_{AA}  (\bm{0} , 0). \label{eq:41}
\end{align}
%and $\lambda_{\rm{EQ}} =(g_{1} + g_{2})  \hspace{1mm} \chi^{\, d,0}_{AA}(\bm{0}, 0)$. 

\begin{figure}[htbp]
  \begin{center}
    \includegraphics[width=0.495\hsize]{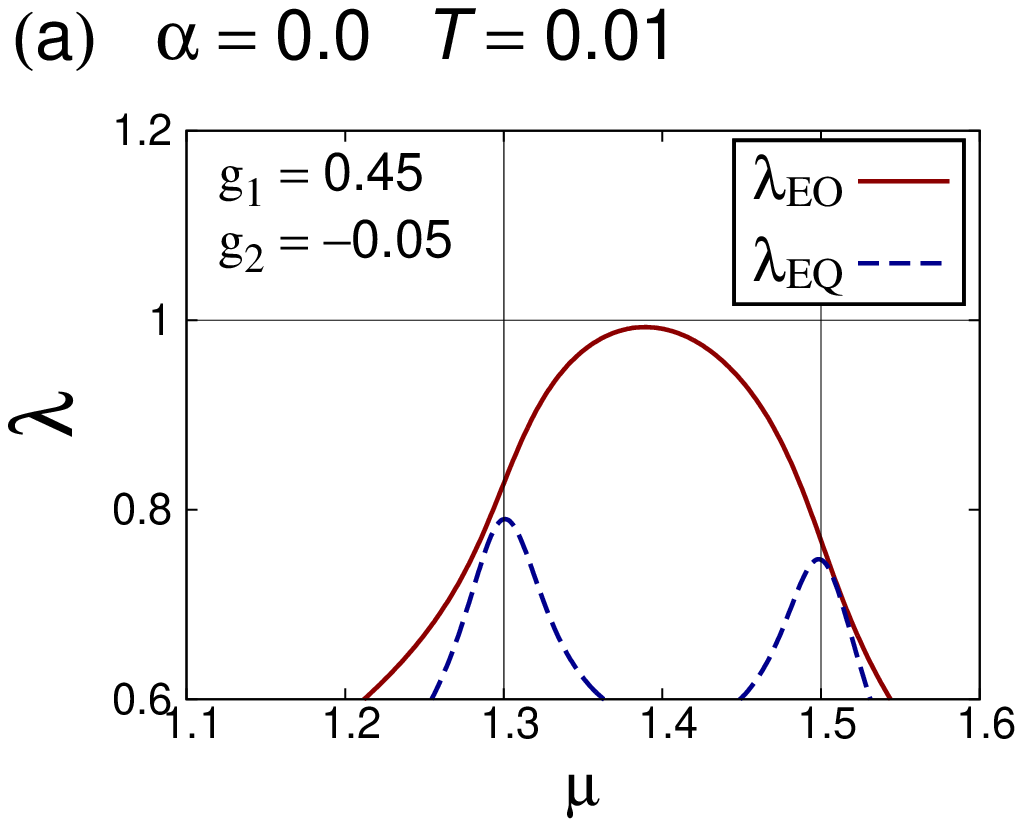}
    \includegraphics[width=0.495\hsize]{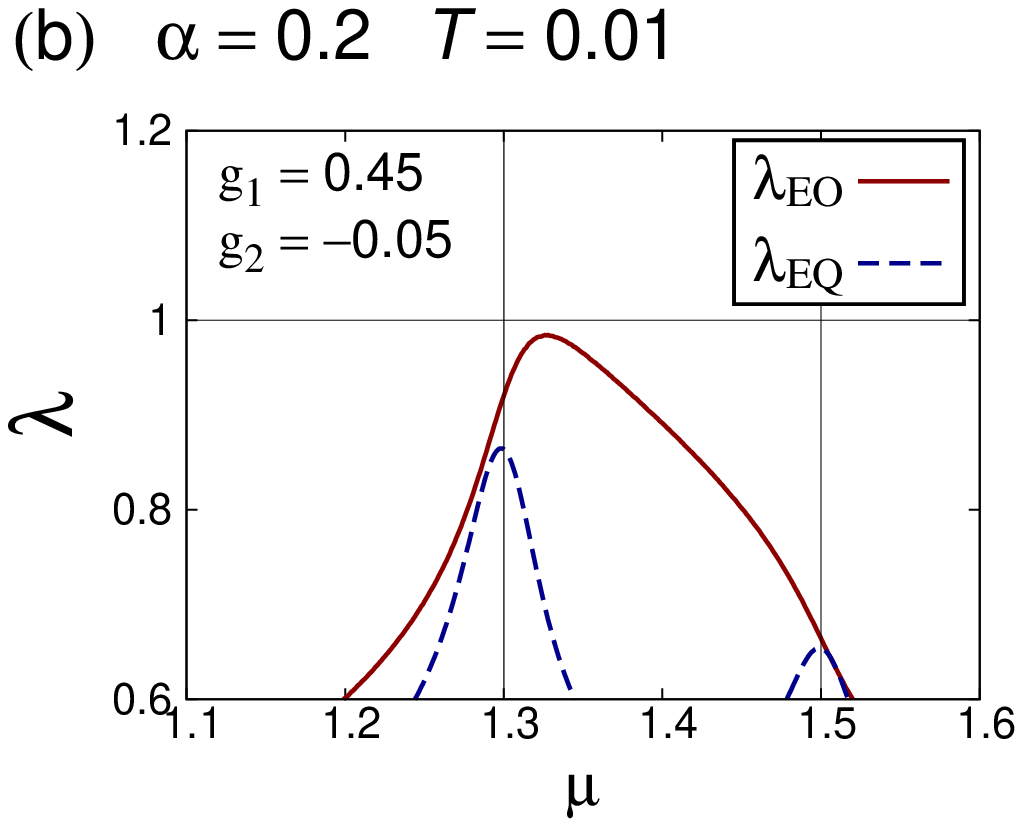}
  \end{center}
  \begin{center}
    \includegraphics[width=0.495\hsize]{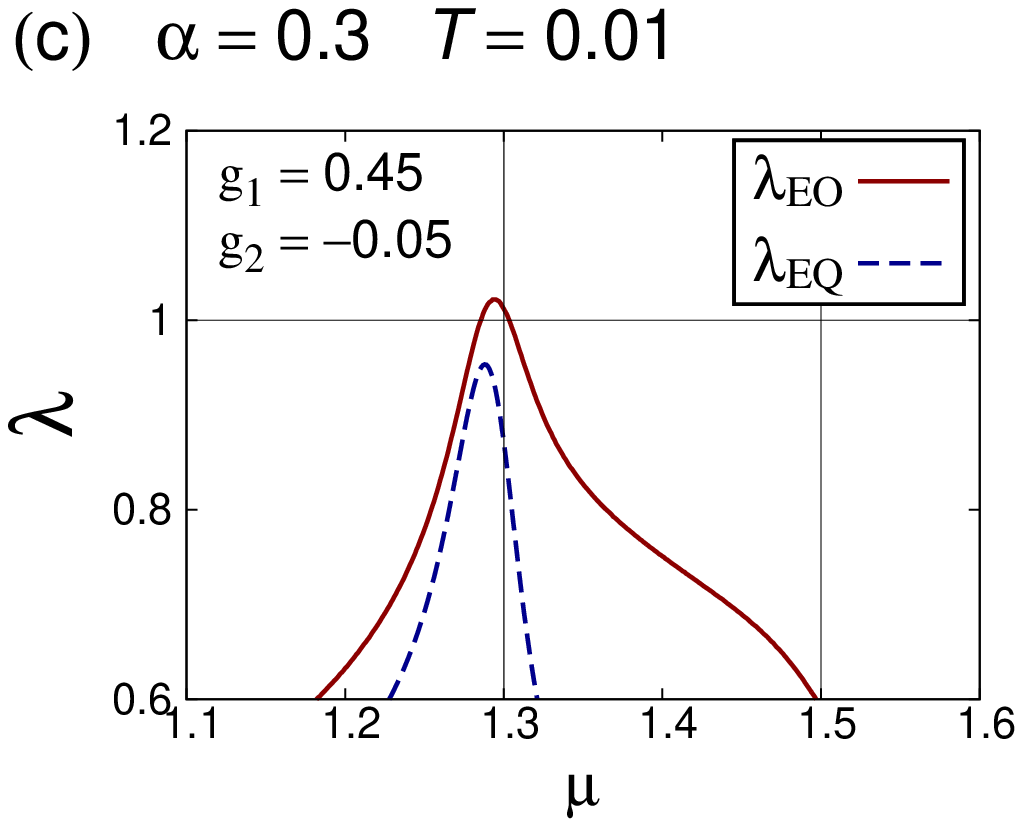}
    \includegraphics[width=0.495\hsize]{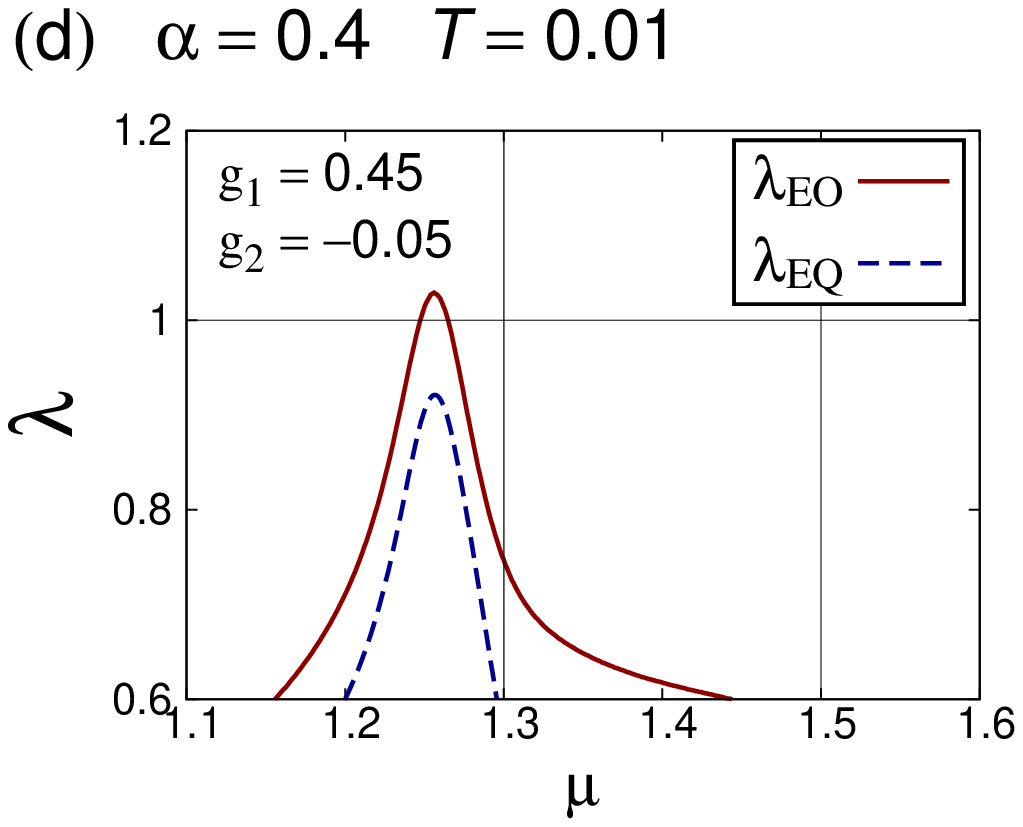}
    \caption{(Color online) Eigenvalues $\lambda_{\rm{EO}} $ and $\lambda_{\rm{EQ}}$ 
at (a) $\alpha = 0.0$, (b) $\alpha = 0.2$, (c) $\alpha = 0.3$, and (d) $\alpha = 0.4$. 
We choose $g_{1} = 0.45$, $g_{2} = -0.05$, and $T=0.01$. 
The red solid lines and blue dashed lines show $\lambda_{\rm{EO}}$ and $\lambda_{\rm{EQ}}$, respectively. 
    }
    \label{fig:chi_T_0.01_soc_dep}
  \end{center}
\end{figure}

Figure~\ref{fig:chi_T_0.01_soc_dep} shows $\lambda_{\rm{EO}}$ and $\lambda_{\rm{EQ}}$.  
Because we assume a rather high temperature $T=0.01$, the intraband contribution is small; thus, 
the EO order is the leading instability even at $\alpha=0$ [Fig.~\ref{fig:chi_T_0.01_soc_dep}(a)]. 
The $\mu$-dependence of $\lambda_{\rm{EO}}$ is quite different from that of $\lambda_{\rm{EQ}}$. 
The peak of $\lambda_{\rm{EO}}$ at around $\mu =1.4$ comes from a large negative interlayer irreducible susceptibility 
$\chi^{\, d,0}_{A B}$. 
The interband contribution is maximized when the chemical potential is in the middle of 
the two van Hove singularities. 
When the ASOC is increased, $\lambda_{\rm{EO}}$ becomes similar to $\lambda_{\rm{EQ}}$
[Figs.~\ref{fig:chi_T_0.01_soc_dep}(c) and \ref{fig:chi_T_0.01_soc_dep}(d)]. 
Since the DOS is increased, the multipole susceptibility is enhanced at around $\mu=1.3$. 
These behaviors are indeed what we expected in Sect.~4.

\begin{figure}[htbp]
  \begin{center}
    \includegraphics[width=0.495\hsize]{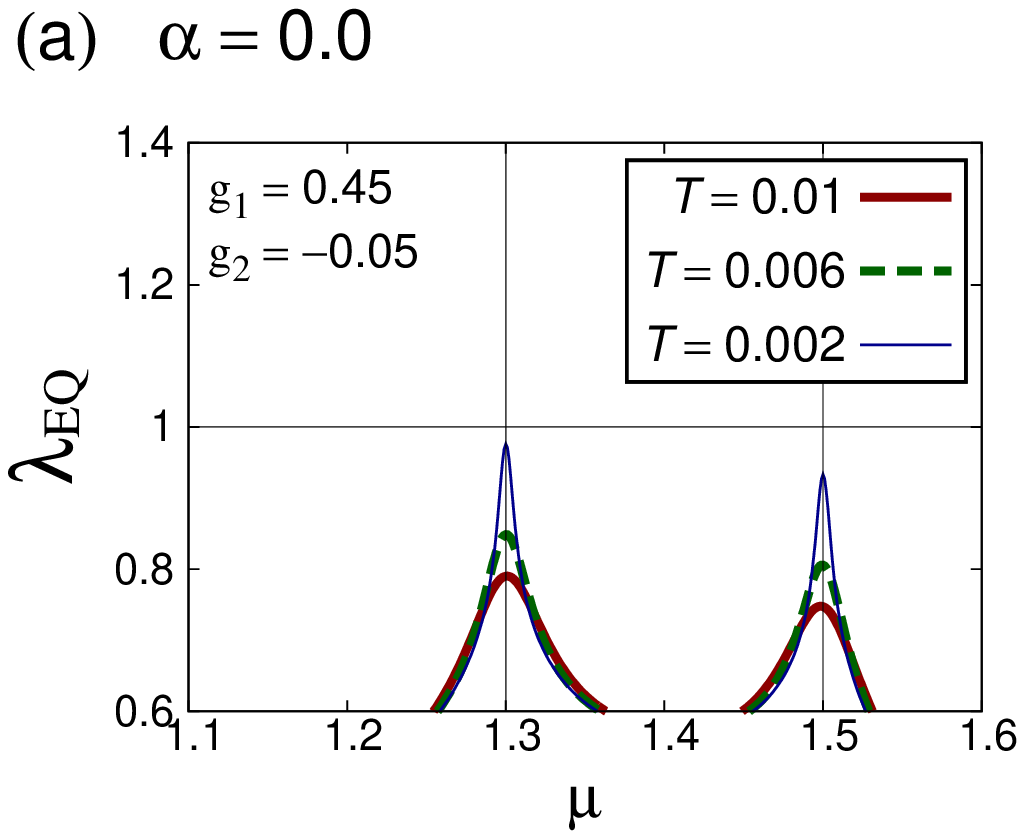}
    \includegraphics[width=0.495\hsize]{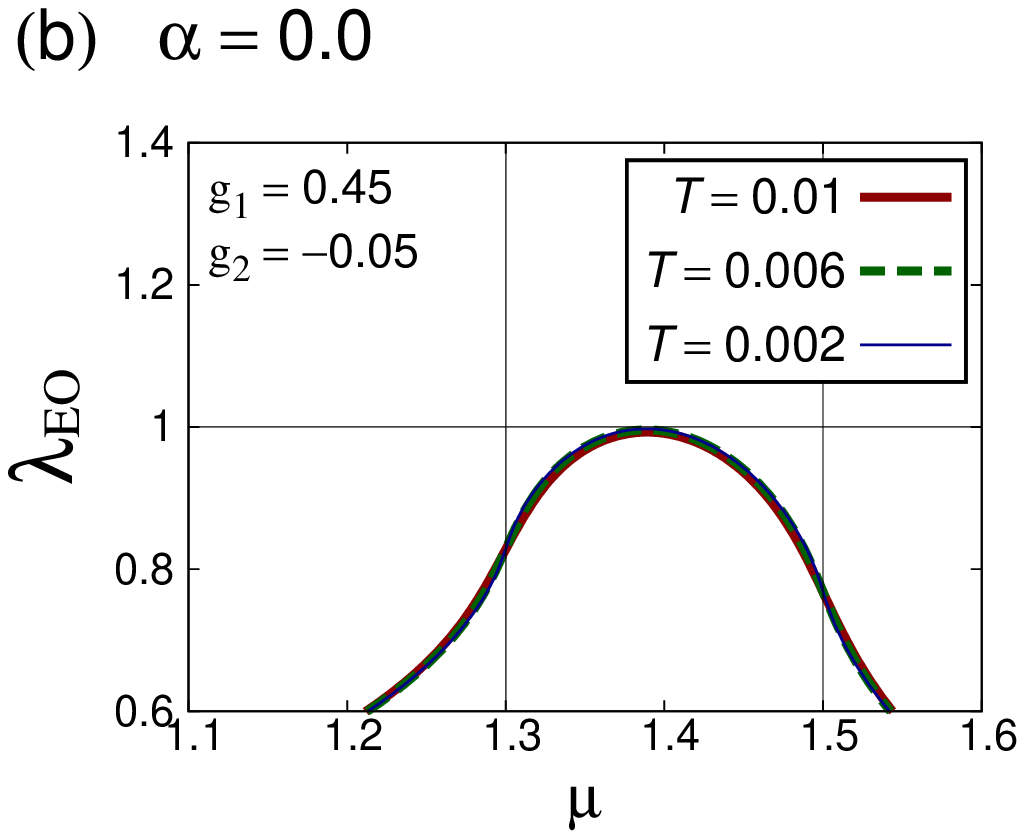}
  \end{center}
  \begin{center}
    \includegraphics[width=0.495\hsize]{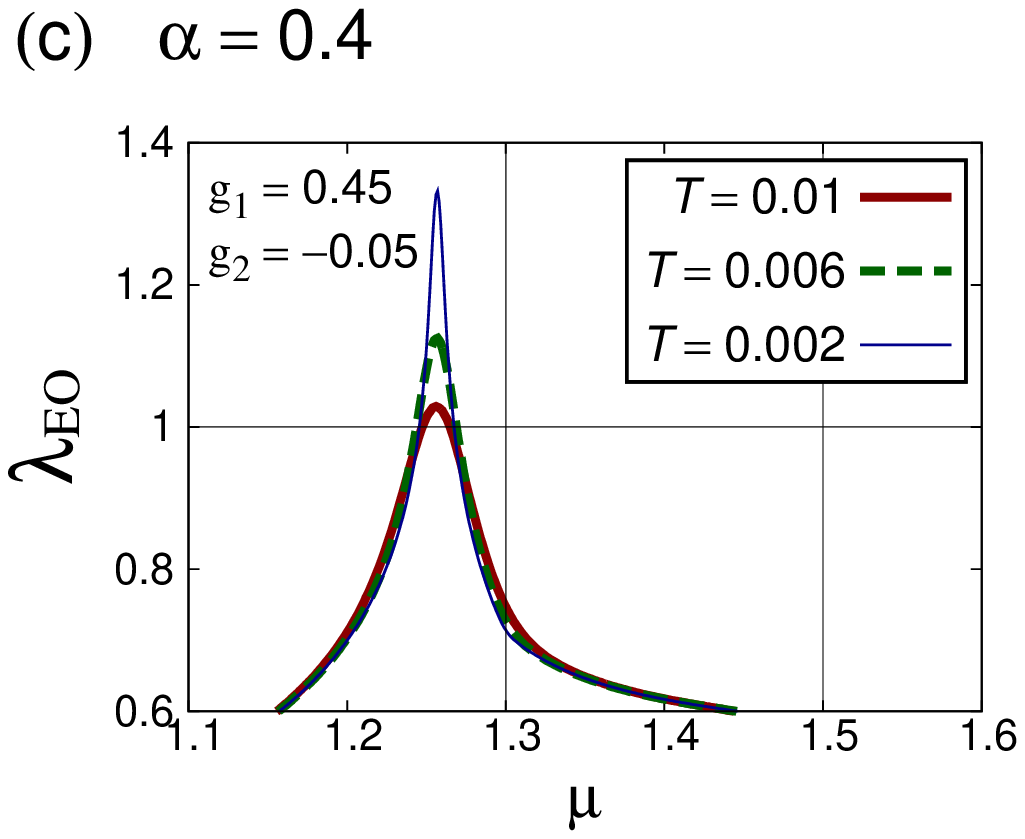}
  \end{center}
  \caption{(Color online) (a) $\lambda_{\rm{EQ}}$ at $\alpha = 0$, (b) $\lambda_{\rm{EO}}$ at $\alpha = 0$, 
and (c) $\lambda_{\rm{EO}}$ at $\alpha = 0.4$ for $g_{1} = 0.45$ and $g_{2} = -0.05$. 
The red thick solid, green dashed, and blue thin solid lines show the results at the temperature 
$T = 0.01$, $0.006$ and $0.002$, respectively.
  }
  \label{fig:lambda_T_dep}
\end{figure}

The Rashba ASOC more significantly affects the temperature dependence of $\lambda_{\rm{EO}}$. 
Figure \ref{fig:lambda_T_dep} shows $\lambda_{\rm{EO}}$ and $\lambda_{\rm{EQ}}$ at three temperatures from $T=0.002$ 
to $T=0.01$. It is shown that the eigenvalue for the EO order $\lambda_{\rm{EO}}$ is almost temperature-independent 
at $\alpha=0$ [Fig.~\ref{fig:lambda_T_dep}(b)], while $\lambda_{\rm{EQ}}$ grows with decreasing temperature 
[Fig.~\ref{fig:lambda_T_dep}(a)]. 
This is because the former comes from the interband contribution and the latter comes from the intraband contribution. 
More specifically, the momentum between the two Fermi surfaces contributes to Eq.~(\ref{eq:38}), while Eq.~(\ref{eq:37}) 
is determined by the momentum in the vicinity of the Fermi surfaces within $|E_{{\bm k}\nu}| < 2T$. 
Therefore, the effect of the large DOS on $\lambda_{\rm{EQ}}$ is smeared by the temperature. 
By decreasing the temperature, the sharp peak of DOS leads to a large $\lambda_{\rm{EQ}}$ at the van Hove singularity. 
This is the reason why the EQ state is favored at low temperatures \cite{Yamase_Bilayer_1,Yamase_Bilayer_2}.

On the other hand, both $\lambda_{\rm{EO}}$ and $\lambda_{\rm{EQ}}$ are obtained by the intraband contribution in the 
presence of the large ASOC, as we showed in Eqs.~(\ref{eq:40}) and (\ref{eq:37}). 
Therefore, $\lambda_{\rm{EO}}$ also grows with decreasing temperature [Fig.~\ref{fig:lambda_T_dep}(c)]. 
We find that the temperature dependence of $\lambda_{\rm{EO}}$ undergoes a significant change from $\alpha = 0.2$ to $0.3$. 
This implies that the electronic structure exhibits a crossover from the bonding and antibonding orbitals to the decoupled layers 
at around $\alpha/t_\perp =2$. This is consistent with studies of multilayer 
superconductors~\cite{Yoshida_PDW_1, Yoshida_PDW_2, Watanabe_PDW,Maruyama_multilayer_SC}.

%%%%%%%%%%%%%%%%%%%%%%%%%%%%%%%%%%%%%%%%%%%%%%%%%%%%%%%%%%%%%%%%%%%
%%%%%%%%%%%%%%%%%%%%%%%%%%%%%%%%%%%%%%%%%%%%%%%%%%%%%%%%%%%%%%%%%%%
%%%%%%%%%%%%%%%%%%%%%%%%%%%%%%%%%%%%%%%%%%%%%%%%%%%%%%%%%%%%%%%%%%%

%%%%%%%%%%%%%%%%%%%%%%%%%%%%%%%%%%%%%%%%%%%%%%%%%%%%%%%%%%%%%%%%%%%
%%%%%%%%%%%%%%%%%%%%%%%%%%%%%%%%%%%%%%%%%%%%%%%%%%%%%%%%%%%%%%%%%%%
%%%%%%%%%%%%%%%%%%%%%%%%%%%%%%%%%%%%%%%%%%%%%%%%%%%%%%%%%%%%%%%%%%%
\subsection{Electronic structure in multipole states}

\begin{figure}[htbp]
  \begin{center}
    \includegraphics[width=0.5\hsize]{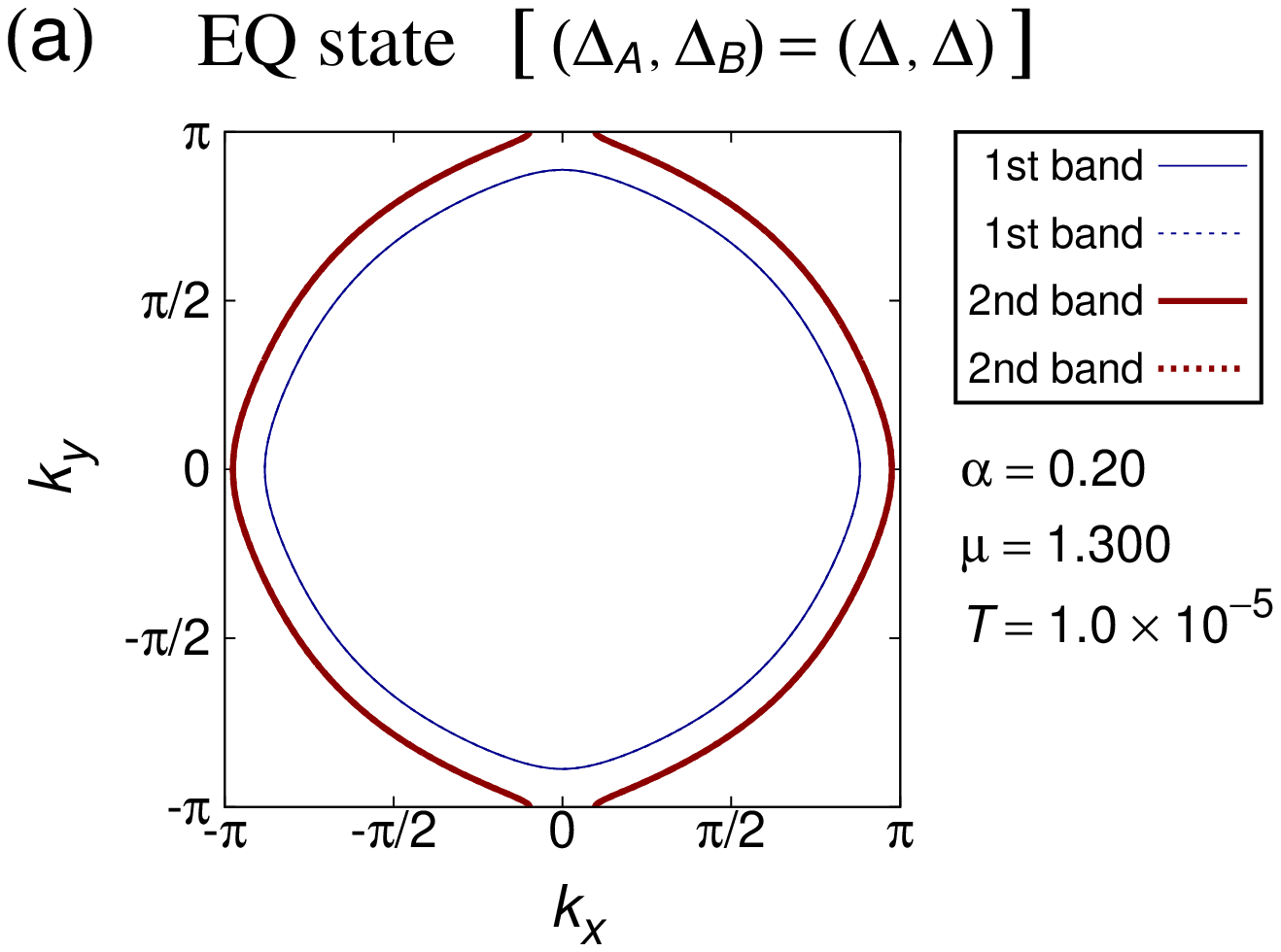}
    \includegraphics[width=0.5\hsize]{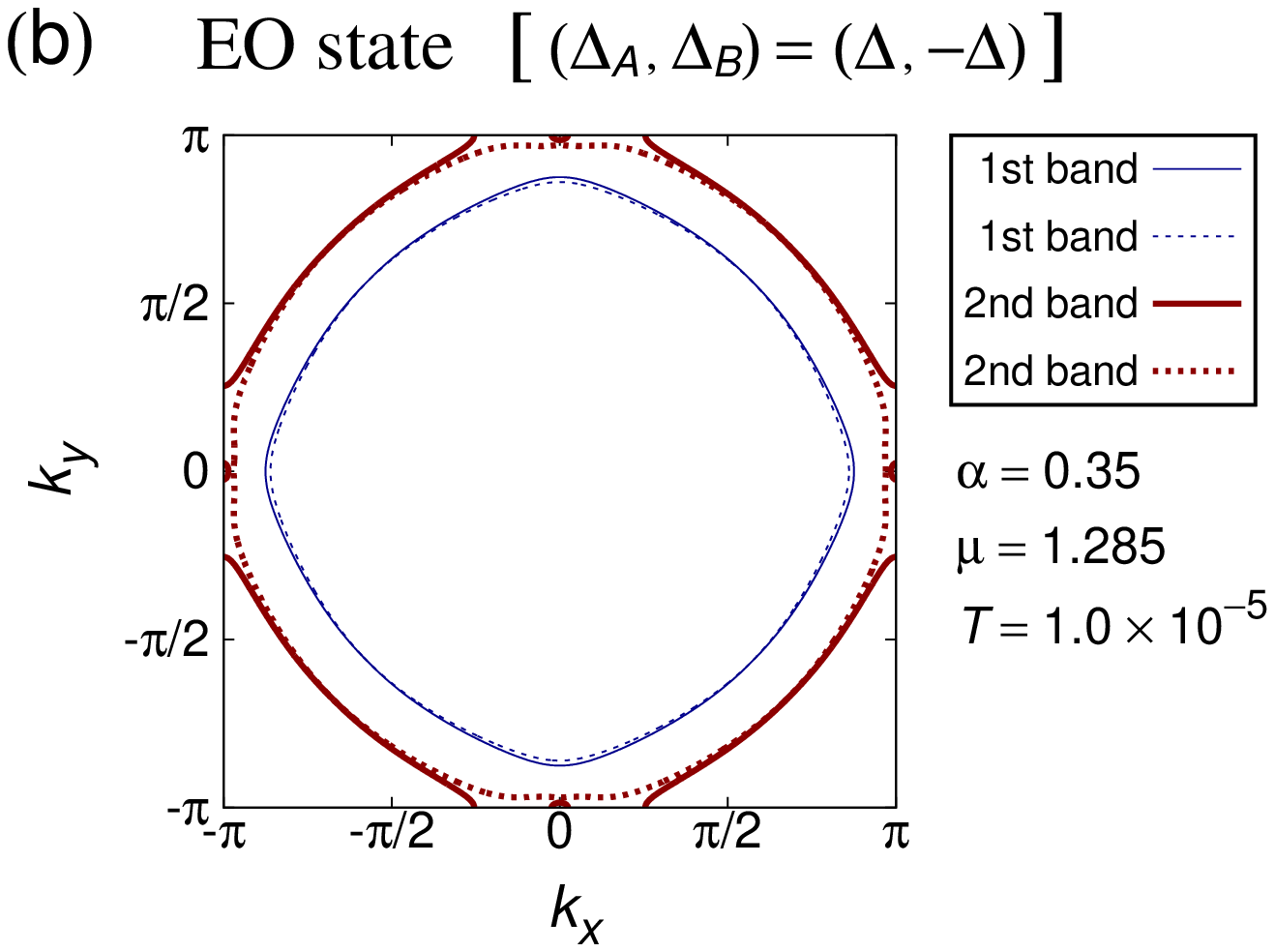}
    \caption{(Color online) Fermi surfaces in (a) EQ and (b) EO states at $g_{1} = 0.45$, $g_{2} = -0.05$, and $T = 1.0 \times 10^{-5}$.
             The blue thin (red thick) lines indicate the Fermi surfaces of the 1st band, $E_{\bm{k} 1}$ and $E_{\bm{k} 2}$ 
             (2nd band, $E_{\bm{k} 3}$ and $E_{\bm{k} 4}$). 
             In (a), we adopt $\alpha = 0.2$ and $\mu = 1.3$ leading to the order parameter 
             $(\Delta_{A}, \Delta_{B}) \simeq ( -0.0025, -0.0025)$.
             In (b), we adopt $\alpha = 0.35$ and $\mu = 1.285$ leading to 
             $(\Delta_{A}, \Delta_{B}) \simeq ( -0.018, 0.018)$. Because of the spontaneous inversion symmetry breaking, 
             $E_{\bm{k} 1} \ne E_{\bm{k} 2}$ and $E_{\bm{k} 3} \ne E_{\bm{k} 4}$ in the EO state. 
    }
    \label{fig:FIG_FS}
 \end{center}
\end{figure}

Now, we demonstrate the electronic structures in the EQ and EO states with particular focus on 
the role of the layer-dependent Rashba ASOC. 
Diagonalizing the mean field Hamiltonian $\hat{H}_{4}^{\, \rm{MF}} (\bm{k})$ in Eq. (\ref{eq:11}), we obtain the eigenvalues 
\begin{subequations}
\begin{align}
E_{\bm{k} 1} &=  \varepsilon_{\bm{k}} + d_{\bm{k}} \frac{\Delta_{A} + \Delta_{B}}{2} + \sqrt{(\alpha^{\, d}_{\bm{k} -})^{2} + t^{2}_{\perp}} , \label{eq:42a} \\
E_{\bm{k} 2} &=  \varepsilon_{\bm{k}} + d_{\bm{k}} \frac{\Delta_{A} + \Delta_{B}}{2} + \sqrt{(\alpha^{\, d}_{\bm{k} +})^{2} + t^{2}_{\perp}} , \label{eq:42b} \\
E_{\bm{k} 3} &=  \varepsilon_{\bm{k}} + d_{\bm{k}} \frac{\Delta_{A} + \Delta_{B}}{2} - \sqrt{(\alpha^{\, d}_{\bm{k} -})^{2} + t^{2}_{\perp}} , \label{eq:42c} \\
E_{\bm{k} 4} &=  \varepsilon_{\bm{k}} + d_{\bm{k}} \frac{\Delta_{A} + \Delta_{B}}{2} - \sqrt{(\alpha^{\, d}_{\bm{k} +})^{2} + t^{2}_{\perp}} , \label{eq:42d}
\end{align}
\end{subequations}
where $\alpha^{\, d}_{\bm{k} \pm}$ is defined as 
\begin{equation}
\alpha^{\, d}_{\bm{k} \pm} \equiv \alpha |\bm{g}_{\bm{k}}| \pm d_{\bm{k}} \frac{\Delta_{A} - \Delta_{B}}{2}. \label{eq:54}
\end{equation}
%\begin{subequations}
%\begin{align}
%E_{\bm{k} 1} &= \varepsilon_{\bm{k}} + d_{\bm{k}} \frac{\Delta_{A} + \Delta_{B}}{2} + \sqrt{ \Biggl( \alpha |\bm{g}_{\bm{k}}| - d_{\bm{k}} \frac{\Delta_{A} - \Delta_{B}}{2} \Biggr)^{2}   + t_{\perp}^{2}  }.  \label{eq:42a} \\
%E_{\bm{k} 2} &= \varepsilon_{\bm{k}} + d_{\bm{k}} \frac{\Delta_{A} + \Delta_{B}}{2} + \sqrt{ \Biggl( \alpha |\bm{g}_{\bm{k}}| + d_{\bm{k}} \frac{\Delta_{A} - \Delta_{B}}{2} \Biggr)^{2}   + t_{\perp}^{2}  }.  \label{eq:42b} \\
%E_{\bm{k} 3} &= \varepsilon_{\bm{k}} + d_{\bm{k}} \frac{\Delta_{A} + \Delta_{B}}{2} - \sqrt{ \Biggl( \alpha |\bm{g}_{\bm{k}}| - d_{\bm{k}} \frac{\Delta_{A} - \Delta_{B}}{2} \Biggr)^{2}   + t_{\perp}^{2}  }.  \label{eq:42c} \\
%E_{\bm{k} 4} &= \varepsilon_{\bm{k}} + d_{\bm{k}} \frac{\Delta_{A} + \Delta_{B}}{2} - \sqrt{ \Biggl( \alpha |\bm{g}_{\bm{k}}| - d_{\bm{k}} \frac{\Delta_{A} - \Delta_{B}}{2} \Biggr)^{2}   + t_{\perp}^{2}  }.  \label{eq:42d}
%\end{align}
%\end{subequations}
%
In the EQ state, $\Delta_{A} = \Delta_{B} = \Delta$, and the dispersion relation is reduced to 
\begin{equation}
E_{\bm{k} \nu} = \varepsilon_{\bm{k}} + d_{\bm{k}} \Delta \pm \sqrt{ \left( \alpha |\bm{g}_{\bm{k}}| \right)^{2}   + t_{\perp}^{2}  }.  \label{eq:42'}
\end{equation}
We see $E_{\bm{k} 1} = E_{\bm{k} 2}$ and $E_{\bm{k} 3} = E_{\bm{k} 4}$. The twofold degeneracy is protected by 
the global inversion symmetry and time-reversal symmetry. 
The layer-dependent Rashba ASOC simply increases the band gap $E_{\bm{k} 1} - E_{\bm{k} 3}$. 
Owing to a small dPI order parameter $(\Delta_{A}, \Delta_{B}) \simeq ( -0.0025, -0.0025)$, 
the Fermi surface is spontaneously deformed, particularly near the van Hove singularities 
[Fig.~\ref{fig:FIG_FS}(a)]. 
This is a characteristic behavior of the nematic state, which has been reported 
in previous studies.~\cite{Yamase_dPI_1, Yamase_dPI_2, Khavkine_dPI, Kee_dPI}.

More interestingly, the twofold degeneracy in the band structure is lifted in the EO state [Fig.~\ref{fig:FIG_FS}(b)]. 
Because the global inversion symmetry is spontaneously broken, the layer-dependent Rashba ASOC lifts the degeneracy. 
For $\Delta_{A} = -\Delta_{B} = \Delta$, the dispersion relation in the EO state is obtained as 
\begin{equation}
E_{\bm{k} \nu} = \varepsilon_{\bm{k}}  \pm \sqrt{ \left( \alpha |\bm{g}_{\bm{k}}| \pm d_{\bm{k}} \Delta \right)^{2}   + t_{\perp}^{2}  }.  \label{eq:42''}
\end{equation}
We see  $E_{\bm{k} 1} \ne E_{\bm{k} 2}$ and $E_{\bm{k} 3} \ne E_{\bm{k} 4}$ when $\alpha \ne 0$. 
This is a characteristic property of the odd-parity electric multipole state~\cite{Hitomi_EO}, 
which is illustrated in Sect. 2.

Next, we discuss the spin texture emerging in the EO state. 
The spin texture of the split bands is characterized by the effective g-vector~\cite{Hitomi_EO,Yanase_multi_orbital_SOC,Nakamura_STO} 
\begin{equation}
\bm{g}^{\, i}_{\bm{k}} = \frac{ E_{\bm{k},2i} - E_{\bm{k},2i-1} }{2} \frac{\bm{S}^{\, \rm{av}}_{\bm{k},2i}}{|\bm{S}^{\, \rm{av}}_{\bm{k},2i}|},  \label{eq:49}
\end{equation}
for the 1st band ($i=1$) and the 2nd band ($i=2$). 
%Considering generic order parameters $(\Delta_{A}, \Delta_{B})$, 
Calculating the expectation values of the spin 
$\bm{S}^{\, \rm{av}}_{\bm{k},\nu} = \langle  \sum_{s,s',l} \bm{\sigma}^{ss'} c^{\dagger}_{\bm{k}sl} c_{\bm{k}s'l} \rangle_{\nu}$ 
for the $\nu$-th eigenstate for generic order parameters $(\Delta_{A}, \Delta_{B})$ (see Appendix B for details), we obtain 
\begin{align}
\bm{g}^{\, 1}_{\bm{k}}   &=  \frac{\sqrt{(\alpha^{\, d}_{\bm{k} +})^{2} + t^{2}_{\perp}} - \sqrt{(\alpha^{\, d}_{\bm{k} -})^{2} + t^{2}_{\perp}}}{2 \sqrt{ \sin^{2} k_{x} + \sin^{2} k_{y}}} \hspace{1mm} \biggl(-\sin k_{y} ,\hspace{0.5mm} \sin k_{x} ,\hspace{0.5mm} 0\biggr), \label{eq:52} \\
\bm{g}^{\, 2}_{\bm{k}}   &= - \bm{g}^{\, 1}_{\bm{k}}.  
%&=  \frac{\sqrt{(\alpha^{\, d}_{\bm{k} +})^{2} + t^{2}_{\perp}} - \sqrt{(\alpha^{\, d}_{\bm{k} -})^{2} + t^{2}_{\perp}}}{2 \sqrt{ \sin^{2} k_{x} + \sin^{2} k_{y}}} \hspace{1mm} \biggl(\sin k_{y} ,\hspace{0.5mm} -\sin k_{x} ,\hspace{0.5mm} 0\biggr), 
\label{eq:53} 
\end{align}
%where $\alpha^{\, d}_{\bm{k} \pm}$ is defined by 
%\begin{equation}
%\alpha^{\, d}_{\bm{k} \pm} \equiv \alpha |\bm{g}_{\bm{k}}| \pm d_{\bm{k}} \frac{\Delta_{A} - \Delta_{B}}{2}, \label{eq:54}
%\end{equation}
Note that $\alpha^{\, d}_{\bm{k} \pm}$ is reduced to $\alpha^{\, d}_{\bm{k} \pm} = \alpha |\bm{g}_{\bm{k}}| \pm d_{\bm{k}} \Delta$ in the EO state.

\begin{figure}[htbp]
 \begin{center}
   \includegraphics[width=0.50\hsize]{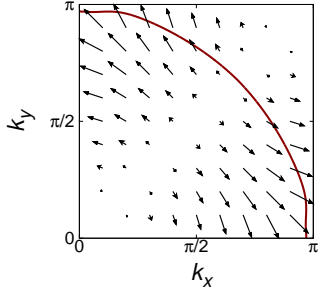}
   \caption{(Color online) Effective g-vector in the 2nd band, $\bm{g}^{\, 2}_{\bm{k}}$. 
   We choose the parameters $g_{1} = 0.45$, $g_{2} = -0.05$, $T = 1.0 \times 10^{-5}$, $\alpha = 0.35$, and $\mu = 1.285$, 
   where the EO state is stabilized. 
   The red solid line shows zeros of $(E_{\bm{k} 3} + E_{\bm{k} 4})/2$. 
   We emphasize the $k_{y} \hat{x} + k_{x} \hat{y}$ symmetry of the g-vector. 
   }
   \label{fig:g_vec}
  \end{center}
\end{figure}

Figure~\ref{fig:g_vec} shows the momentum dependence of the effective g-vector $\bm{g}^{\, 2}_{\bm{k}}$ 
in the quarter Brillouin zone. 
The length of the arrow expresses the spin splitting energy, and the direction represents the spin texture. 
Notice that the spin texture is not of the Rashba type. The symmetry of the spin texture 
is $k_{y} \hat{x} + k_{x} \hat{y}$, which is characteristic of the $D_{2d}$ point group symmetry. 
Indeed, the symmetry of the EO state is not the $C_{4v}$ point group leading to the Rashba-type spin texture~\cite{Springer_NCS}, 
but the $D_{2d}$ point group.~\cite{Hitomi_EO}

Now, we prove the symmetry of the spin texture. 
The generators of the $D_{2d}$ point group are mirror reflection with respect to the $xz$-plane ($M_{xz}$) 
and $\pi/2$ rotation along the $z$-axis combined with the mirror reflection with respect to the $xy$-plane ($S_4$). 
By $M_{xz}$, the momentum and spin are transformed as 
\begin{align}
(k_{x}, k_{y}, k_{z}) & \rightarrow (k_{x}, -k_{y}, k_{z}),  \label{eq:55} \\
(s_{x}, s_{y}, s_{z}) & \rightarrow (-s_{x}, s_{y}, -s_{z}), \label{eq:56}
\end{align}
while, by $S_4$, we have  
\begin{align}
& (k_{x}, k_{y}, k_{z})   \xrightarrow[ z-\rm{axis} ]{ \pi /2 \hspace{0.5mm} \rm{Rot.} } (-k_{y},  k_{x}, k_{z}) \xrightarrow[ xy-\rm{plane} ]{ \rm{Mir.} \hspace{0.5mm} \rm{ref.}} (-k_{y}, k_{x}, -k_{z}), \label{eq:57} \\
& (s_{x}, s_{y}, s_{z})   \rightarrow (-s_{y},  s_{x}, s_{z}) \rightarrow (s_{y}, -s_{x}, s_{z}). \label{eq:58}
\end{align}
The ASOC represented by the g-vector with the $k_{y} \hat{x} + k_{x} \hat{y}$ symmetry is invariant 
under all the symmetry operations of the $D_{2d}$ point group~\cite{Comment1}.

%%%%%%%%%%%%%%%%%%%%%%%%%%%%%%%%%%%%%%%%%%%%%%%%%%%%%%%%%%%%%%%%%%%
%%%%%%%%%%%%%%%%%%%%%%%%%%%%%%%%%%%%%%%%%%%%%%%%%%%%%%%%%%%%%%%%%%%
%%%%%%%%%%%%%%%%%%%%%%%%%%%%%%%%%%%%%%%%%%%%%%%%%%%%%%%%%%%%%%%%%%%

%%%%%%%%%%%%%%%%%%%%%%%%%%%%%%%%%%%%%%%%%%%%%%%%%%%%%%%%%%%%%%%%%%%
%%%%%%%%%%%%%%%%%%%%%%%%%%%%%%%%%%%%%%%%%%%%%%%%%%%%%%%%%%%%%%%%%%%
%%%%%%%%%%%%%%%%%%%%%%%%%%%%%%%%%%%%%%%%%%%%%%%%%%%%%%%%%%%%%%%%%%%
\section{Electric Octupole State in Magnetic Field}

Finally, we investigate the EO state under the magnetic field. 
Stimulated by experimental indications of nematic order in Sr$_3$Ru$_2$O$_7$~\cite{Borzi,Stingl,Mackenzie}, 
the magnetic-field-induced dPI order has been studied theoretically~\cite{Kee_dPI,Yamase-Katanin,Yamase2007,Yamase_Bilayer_1}. 
Then, it has been shown that the dPI order may be caused by the magnetic field tuning the Fermi surface 
to be close to the van Hove singularity. 
Here we examine the phase diagram of the bilayer Rashba model in the magnetic field.

\begin{figure}[htbp]
 \begin{center}
   \includegraphics[width=0.6 \hsize]{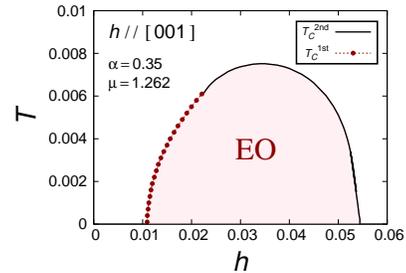}
   \caption{(Color online) $h$-$T$ phase diagram in the magnetic field along the $[001]$-axis. 
                           We choose the parameters $g_{1} = 0.45$, $g_{2} = -0.05$, $\alpha = 0.35$, and $\mu = 1.262$.
                           The red circles indicate the first-order transition line $T_{\rm c}^{\rm 1st}$, while 
                           the black line shows the second-order transition line $T_{\rm c}^{\rm 2nd}$.
                           In the shaded region, the EO state is stabilized.
   }
   \label{fig:FIG_h_T}
 \end{center}
\end{figure}

In order to take into account the magnetic field, we add the Zeeman coupling term  
\begin{equation}
H_{\rm{Zeeman}} = - \sum_{\bm{k},s,s',l} \bm{h} \cdot \bm{\sigma}^{ss'} c_{\bm{k}sl}^{\dagger} \, c_{\bm{k}s'l} ,   \label{eq:59} \\
\end{equation}
to the Hamiltonian $H$. 
Figure \ref{fig:FIG_h_T} shows the obtained $h$-$T$ phase diagram for $\bm{h} = h \, \hat{z} \parallel$ [001] in the large ASOC region.
The dome-shaped transition temperature is shown. 
We find that the ordered phase is occupied by the EO phase.
Interestingly, the order of quantum phase transition is different from that in the monolayer and bilayer forward scattering models 
in the absence of the ASOC~\cite{Kee_dPI,Yamase-Katanin,Yamase2007,Yamase_Bilayer_1}. 
Although the quantum dPI transition is of the first order at $\alpha=0$, the quantum critical point 
emerges in our bilayer Rashba model at $h=0.054$. Therefore, the bilayer Rashba model may be 
a platform for the nematic quantum criticality~\cite{Oganesyan,Metzner2003,Metlitski,Holder} 
or parity-breaking quantum criticality.

\begin{figure}[htbp]
  \begin{center}
    \includegraphics[width=0.5 \hsize]{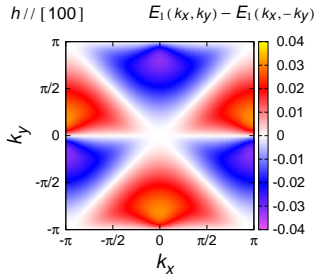}
    \caption{(Color online) Asymmetry of the band structure, $E_{\rm{1}} (k_{x} , k_{y}) - E_{\rm{1}} (k_{x} , -k_{y})$, 
             in the magnetic field along the [100]-axis, ${\bm h} = h \hat{x}$. 
             We choose the parameters $g_{1} = 0.45$, $g_{2} = -0.05$, $\alpha = 0.35$, $\mu = 1.262$, 
             $T = 1.0 \times 10^{-6}$, and $h = 0.02$. 
    }
    \label{fig:FIG_Ey_01th}
  \end{center}
\end{figure}

We illustrate an intriguing property of the EO state under the in-plane magnetic field. 
Adopting the parameters $g_{1} = 0.45$, $g_{2} = -0.05$, $\alpha = 0.35$, $\mu = 1.262$, $T = 1.0 \times 10^{-6}$, 
and ${\bm h} = 0.02 \, \hat{x}$, we obtain the order parameter $(\Delta_{A}, \Delta_{B}) \simeq ( -0.0097, 0.0097)$. 
Figure~\ref{fig:FIG_Ey_01th} plots $E_{\rm{1}} (k_{x} , k_{y}) - E_{\rm{1}} (k_{x} , -k_{y})$, 
where $E_{\nu} (k_{x} , k_{y})$ is obtained by ordering $E_{\bm{k} \nu}$ as 
$E_{1} (k_{x} , k_{y}) \geq E_{2} (k_{x} , k_{y}) \geq E_{3} (k_{x} , k_{y}) \geq E_{4} (k_{x} , k_{y})$. 
%$E_{\bm{k} \nu}$ is represented by $E_{\nu} (k_{x} , k_{y})$.
This quantity indicates the asymmetry in the band structure. 
Thus, Fig.~\ref{fig:FIG_Ey_01th} shows the asymmetric band structure in the EO state, 
which is analogous to the band shift due to the Zeeman coupling term in noncentrosymmetric metals~\cite{Springer_NCS}.

In order to clarify the asymmetric band structure, we introduce the effective two-band model,  
following Refs.~\citen{Hitomi_EO, Yanase_multi_orbital_SOC}, and \citen{Nakamura_STO}, 
\begin{equation}
H_{\rm{eff}} = \sum^{2}_{i=1} \sum_{\bm{k},s,s'} \Bigl[ \xi^{\, i}_{\bm{k}} \sigma^{ss'}_{0} + ( \bm{g}^{\, i}_{\bm{k}} - \bm{h} ) \cdot \bm{\sigma}^{ss'} \Bigr] \hspace{1mm} a^{\dagger}_{\bm{k} s i} \, a_{\bm{k} s' i}, \label{eq:60}
\end{equation}
with $\xi^{\, i}_{\bm{k}} = (E_{\bm{k},2i} + E_{\bm{k},2i-1})/2$. 
Assuming a Zeeman energy much smaller than the magnitude of ASOC, we obtain the dispersion relation 
\begin{equation}
E_{\bm{k},\pm, i} = \xi^{\, i}_{\bm{k}} \pm |\bm{g}^{\, i}_{\bm{k}} - \bm{h}| \simeq  \xi^{\, i}_{\bm{k}} \pm |\bm{g}^{\, i}_{\bm{k}}| \mp \frac{\bm{g}^{\, i}_{\bm{k}} \cdot \bm{h}}{|\bm{g}^{\, i}_{\bm{k}}|}. \label{eq:61}
\end{equation}
The asymmetry arises from the last term $\mp (\bm{g}^{\, i}_{\bm{k}} \cdot \bm{h}) / |\bm{g}^{\, i}_{\bm{k}}|$. 
Now, we understand the magnetic field angle dependence of the asymmetry. 
In the EO state, the effective g-vector is obtained in Eqs. (\ref{eq:52}) and (\ref{eq:53}) as 
$\bm{g}^{\, i}_{\bm{k}} \propto (\sin k_{y} , \sin k_{x} , 0)$. 
Thus, we obtain the asymmetry in the band structure 
\begin{align}
E_{\nu} (k_{x} , k_{y}) &= E_{\nu} (-k_{x} , k_{y}) \neq E_{\nu} (k_{x} , -k_{y}) \hspace{4mm} {\rm for} \hspace{2mm} {\bm h} \parallel [100], \label{eq:62} \\
E_{\nu} (k_{x} , k_{y}) &= E_{\nu} (k_{x} , -k_{y}) \neq E_{\nu} (-k_{x} , k_{y}) \hspace{4mm} {\rm for} \hspace{2mm} {\bm h} \parallel [010]. \label{eq:63}
\end{align}

%%%%%%%%%%%%%%%%%%%%%%%%%%%%%%%%%%%%%%%%%%%%%%%%%%%%%%%%%%%%%%%%%%%
%%%%%%%%%%%%%%%%%%%%%%%%%%%%%%%%%%%%%%%%%%%%%%%%%%%%%%%%%%%%%%%%%%%
%%%%%%%%%%%%%%%%%%%%%%%%%%%%%%%%%%%%%%%%%%%%%%%%%%%%%%%%%%%%%%%%%%%
%%%%%%%%%%%%%%%%%%%%%%%%%%%%%%%%%%%%%%%%%%%%%%%%%%%%%%%%%%%%%%%%%%%
%%%%%%%%%%%%%%%%%%%%%%%%%%%%%%%%%%%%%%%%%%%%%%%%%%%%%%%%%%%%%%%%%%%

%%%%%%%%%%%%%%%%%%%%%%%%%%%%%%%%%%%%%%%%%%%%%%%%%%%%%%%%%%%%%%%%%%%
%%%%%%%%%%%%%%%%%%%%%%%%%%%%%%%%%%%%%%%%%%%%%%%%%%%%%%%%%%%%%%%%%%%
%%%%%%%%%%%%%%%%%%%%%%%%%%%%%%%%%%%%%%%%%%%%%%%%%%%%%%%%%%%%%%%%%%%
%%%%%%%%%%%%%%%%%%%%%%%%%%%%%%%%%%%%%%%%%%%%%%%%%%%%%%%%%%%%%%%%%%%
%%%%%%%%%%%%%%%%%%%%%%%%%%%%%%%%%%%%%%%%%%%%%%%%%%%%%%%%%%%%%%%%%%%
\section{Summary and Discussion}

In this work, we investigate multipole order in the bilayer Rashba system 
on the basis of the forward scattering model. 
We find that the odd-parity EO state is stabilized by the layer-dependent Rashba ASOC in a certain parameter region, 
although the even-parity EQ state is stable in the absence of the ASOC. 

The mechanism of the spin-orbit-coupling-induced EO order is clarified by analytically and numerically 
calculating the multipole susceptibilities. 
A large ASOC decreases the kinetic energy due to the bilayer coupling, because the quasiparticle wave function 
is localized on a layer~\cite{Maruyama_multilayer_SC}. Then, the gain of kinetic energy in the EQ state is suppressed, 
and the EO order may be stabilized particularly in the presence of a repulsive interlayer forward scattering interaction. 
This mechanism for the odd-parity multipole order is analogous to the origin of the odd-parity superconductivity 
in multilayer Rashba superconductors~\cite{Yoshida_PDW_1}. In the latter, the odd-parity pair-density-wave state is stabilized 
by the layer-dependent Rashba ASOC, which suppresses the interlayer Josephson coupling.

The electronic structure in the multipole states has been elucidated as follows.
The $C_4$ rotation symmetry of the band structure is broken in the EQ state. 
More interestingly, the spin splitting occurs in the band structure of the EO state. 
This is a consequence of the spontaneous inversion symmetry breaking. 
The spin texture shows the $k_{y} \hat{x} + k_{x} \hat{y}$ momentum dependence compatible with 
the $D_{2d}$ point group symmetry. 
The spin texture leads to the asymmetric band structure in the magnetic field along the 2D conducting plane. 
When the superconductivity occurs in the asymmetric band, the helical superconducting state 
is stable~\cite{Springer_NCS}. 
The simultaneous violation of inversion symmetry and time-reversal symmetry causes such exotic superconducting state. 
Indeed, a recent study demonstrated the helical superconductivity in the odd-parity magnetic multipole state~\cite{Sumita}. 
Then, the external magnetic field is not needed because the magnetic multipole order spontaneously breaks both 
inversion and time-reversal symmetries. 

We also show that the magnetic field along the $[001]$-axis may stabilize the EO state. 
The $h$-$T$ phase diagram shows a dome shape. 
Interestingly, the quantum critical point between the inversion symmetric state and the asymmetric state appears 
at high magnetic fields. Thus, the parity-breaking quantum criticality may be realized in bilayer systems.

The forward scattering model was originally derived from a 2D Hubbard model~\cite{Halboth_Metzner_dPI_RG_1, Halboth_Metzner_dPI_RG_2,Honerkamp_dPI_RG, Metzner_dPI_RG}, and discussed in order to uncover anomalous properties in high-$T_{\rm c}$ 
cuprate superconductors. 
Recently, Matsuda {\it el al.} have detected a signature of phase transition in the bilayer 
YBa$_2$Cu$_3$O$_{7-x}$~\cite{Matsuda_private} at a temperature higher than the onset temperature of 
charge density wave order~\cite{CDW1,CDW2}. 
The observed change in the nematicity~\cite{Comment2} suggests a symmetry breaking related to the nematic order. 
However, the nematic EQ order is unlikely because the strong $C_4$ rotation symmetry breaking has not been observed so far. 
On the other hand, the EO order breaks inversion symmetry, and may be associated with a sizable change in the nematicity, 
consistent with experimental indications. 
More importantly, the broken inversion symmetry is consistent with the observed anomalous linear dichroism~\cite{Lubashevsky}. 
Furthermore, the EO order may be a source of the crisscrossed stripe order~\cite{Maharaj}, 
which may explain the anomalous features of linear dichroism~\cite{Lubashevsky} 
and charge density wave correlations~\cite{CDW_q_dependence} in a coherent way. 
Under the $D_{\rm 2d}$ point group symmetry of the EO state, the stripe order naturally forms the crisscrossed structure. 
In the future, it is desirable to examine the EO order in bilayer high-$T_{\rm c}$ cuprate superconductors.

%%%%%%%%%%%%%%%%%%%%%%%%%%%%%%%%%%%%%%%%%%%%%%%%%%%%%%%%%%%%%%%%%%%
%%%%%%%%%%%%%%%%%%%%%%%%%%%%%%%%%%%%%%%%%%%%%%%%%%%%%%%%%%%%%%%%%%%
%%%%%%%%%%%%%%%%%%%%%%%%%%%%%%%%%%%%%%%%%%%%%%%%%%%%%%%%%%%%%%%%%%%
%%%%%%%%%%%%%%%%%%%%%%%%%%%%%%%%%%%%%%%%%%%%%%%%%%%%%%%%%%%%%%%%%%%
%%%%%%%%%%%%%%%%%%%%%%%%%%%%%%%%%%%%%%%%%%%%%%%%%%%%%%%%%%%%%%%%%%%

%%%%%%%%%%%%%%%%%%%%%%%%%%%%%%%%%%%%%%%%%%%%%%%%%%%%%%%%%%%%%%%%%%%
%%%%%%%%%%%%%%%%%%%%%%%%%%%%%%%%%%%%%%%%%%%%%%%%%%%%%%%%%%%%%%%%%%%
%%%%%%%%%%%%%%%%%%%%%%%%%%%%%%%%%%%%%%%%%%%%%%%%%%%%%%%%%%%%%%%%%%%
%%%%%%%%%%%%%%%%%%%%%%%%%%%%%%%%%%%%%%%%%%%%%%%%%%%%%%%%%%%%%%%%%%%
%%%%%%%%%%%%%%%%%%%%%%%%%%%%%%%%%%%%%%%%%%%%%%%%%%%%%%%%%%%%%%%%%%%
\section*{Acknowledgements}
The authors are grateful to Y. Matsuda and R. Shiina for fruitful discussions.
Part of the numerical computation in this work was carried out at the Yukawa Institute Computer Facility. 
This work was supported by a Grant-in Aid for Scientific Research on Innovative Areas 
``J-Physics'' (JP15H05884) and ``Topological Materials Science'' (JP16H00991)
from JSPS of Japan, and by JSPS KAKENHI Grants (Numbers JP15K05164 and JP15H05745). 
T. H. is supported by a JSPS Fellowship for Young Scientists. 
%%%%%%%%%%%%%%%%%%%%%%%%%%%%%%%%%%%%%%%%%%%%%%%%%%%%%%%%%%%%%%%%%%%
%%%%%%%%%%%%%%%%%%%%%%%%%%%%%%%%%%%%%%%%%%%%%%%%%%%%%%%%%%%%%%%%%%%
%%%%%%%%%%%%%%%%%%%%%%%%%%%%%%%%%%%%%%%%%%%%%%%%%%%%%%%%%%%%%%%%%%%
%%%%%%%%%%%%%%%%%%%%%%%%%%%%%%%%%%%%%%%%%%%%%%%%%%%%%%%%%%%%%%%%%%%
%%%%%%%%%%%%%%%%%%%%%%%%%%%%%%%%%%%%%%%%%%%%%%%%%%%%%%%%%%%%%%%%%%%

%%%%%%%%%%%%%%%%%%%%%%%%%%%%%%%%%%%%%%%%%%%%%%%%%%%%%%%%%%%%%%%%%%%
%%%%%%%%%%%%%%%%%%%%%%%%%%%%%%%%%%%%%%%%%%%%%%%%%%%%%%%%%%%%%%%%%%%
%%%%%%%%%%%%%%%%%%%%%%%%%%%%%%%%%%%%%%%%%%%%%%%%%%%%%%%%%%%%%%%%%%%
%%%%%%%%%%%%%%%%%%%%%%%%%%%%%%%%%%%%%%%%%%%%%%%%%%%%%%%%%%%%%%%%%%%
%%%%%%%%%%%%%%%%%%%%%%%%%%%%%%%%%%%%%%%%%%%%%%%%%%%%%%%%%%%%%%%%%%%
\appendix
\section{Irreducible Susceptibility [Eqs.~(\ref{eq:27}) and (\ref{eq:28})]}
We show the derivation of irreducible susceptibility in Eqs.~(\ref{eq:27}) and (\ref{eq:28}). 
We start with the generic form Eq. (\ref{eq:16}) and calculate the coefficients at ${\bm q}=0$ 
\begin{equation}
A^{\nu \nu'}_{\bm{k},sl,s'l'} (\bm{0})  =  u^{\nu}_{\bm{k} s l} u^{\nu *}_{\bm{k} s' l'} u^{\nu'}_{\bm{k} s' l'} u^{\nu' *}_{\bm{k} s l} \, , \label{eq:A_1}
\end{equation}
where $u^{\nu}_{\bm{k} s l}$ is given by the unitary matrix diagonalizing the noninteracting Hamiltonian 
$H_{0} = H_{\rm{kin}} + H_{\rm{ASOC}} + H_{\perp}$. 
As in Eq.~(\ref{eq:9}), $H_0$ is represented by the $4 \times 4$ matrix $\hat{H}_{0} (\bm{k})$, i.e.,   
$H_{0} = \sum_{\bm{k}} \hat{C}^{\dagger}_{\bm{k}} \hat{H}_{0} (\bm{k}) \hat{C}_{\bm{k}}$, with 
\begin{equation}
\hat{H}_{0} (\bm{k}) =
\scalebox{1.2}{$\displaystyle
{\footnotesize
\begin{pmatrix}
\varepsilon_{\bm{k}}          & -\alpha \lambda_{\bm{k}}^{+}      & t_{\perp}                          & 0 \\
-\alpha \lambda_{\bm{k}}^{-}   & \varepsilon_{\bm{k}}             & 0                                & t_{\perp} \\
t_{\perp}                     & 0                              & \varepsilon_{\bm{k}}               & \alpha \lambda_{\bm{k}}^{+} \\
0                           & t_{\perp}                        &  \alpha \lambda_{\bm{k}}^{-}        & \varepsilon_{\bm{k}}
\end{pmatrix}. 
} $}
\label{eq:A_2}
\end{equation}
Diagonalizing $\hat{H}_{0} (\bm{k})$ using the unitary matrix $\hat{U} (\bm{k})$ 
\begin{equation}
\hat{U}^{\dagger} (\bm{k}) \hat{H}_{0} (\bm{k}) \hat{U} (\bm{k}) =
\scalebox{1.2}{$\displaystyle
{\footnotesize
\begin{pmatrix}
E_{\bm{k} 1}   &  0            &  0             & 0 \\
0           &  E_{\bm{k} 2}    &  0             & 0 \\
0           &  0             &  E_{\bm{k} 3}    & 0 \\
0           &  0             &  0             & E_{\bm{k} 4}
\end{pmatrix} ,
} $}
\label{eq:A_3}
\end{equation}
we obtain the $\nu$-th eigenstate with the dispersion relation in Eqs.~(\ref{eq:29}) and (\ref{eq:30}). 
The unitary matrix is obtained as
\begin{align}
\hat{U} & (\bm{k}) = \frac{1}{\sqrt{2}} \times \notag \\
&
\scalebox{1.1}{$\displaystyle
{\footnotesize
\begin{pmatrix}
T_{\bm{k}}  &  -\frac{\lambda_{\bm{k}}^{+}}{|\bm{g}_{\bm{k}}|} \sqrt{1 - T_{\bm{k}}^{2}}  &  \sqrt{1 - T_{\bm{k}}^{2}}  &  -\frac{\lambda_{\bm{k}}^{+}}{|\bm{g}_{\bm{k}}|} T_{\bm{k}} \\
\\
\frac{\lambda_{\bm{k}}^{-}}{|\bm{g}_{\bm{k}}|} T_{\bm{k}}  &  \sqrt{1 - T_{\bm{k}}^{2}}  &  \frac{\lambda_{\bm{k}}^{-}}{|\bm{g}_{\bm{k}}|} \sqrt{1 - T_{\bm{k}}^{2}}  &  T_{\bm{k}} \\
\\
\sqrt{1 - T_{\bm{k}}^{2}}  &  -\frac{\lambda_{\bm{k}}^{+}}{|\bm{g}_{\bm{k}}|} T_{\bm{k}}  &  -T_{\bm{k}}  &  \frac{\lambda_{\bm{k}}^{+}}{|\bm{g}_{\bm{k}}|}  \sqrt{1 - T_{\bm{k}}^{2}} \\
\\
\frac{\lambda_{\bm{k}}^{-}}{|\bm{g}_{\bm{k}}|}  \sqrt{1 - T_{\bm{k}}^{2}}  &  T_{\bm{k}}  &  -\frac{\lambda_{\bm{k}}^{-}}{|\bm{g}_{\bm{k}}|} T_{\bm{k}}  &  -\sqrt{1 - T_{\bm{k}}^{2}}
\end{pmatrix} ,
} $}
\label{eq:A_6}
\end{align}
where $T_{\bm{k}}$ is defined in Eq.~(\ref{eq:31}).

In order to show a compact form of $A^{\nu \nu'}_{\bm{k}, sl, s'l'} (\bm{0})$, we introduce the indices 
$\omega_{1}$ and $\omega_{2}$ in terms of $\nu$, 
\begin{equation}
\omega_{1} =
   \begin{cases}
     1  \hspace{1.4cm}  (\nu = 1 , 2),  \\
     2  \hspace{1.4cm}  (\nu = 3 , 4),
   \end{cases} \label{eq:A_8}
\end{equation}
\begin{equation}
\omega_{2} =
   \begin{cases}
     1  \hspace{1.4cm}  (\nu = 1 , 4),  \\
     2  \hspace{1.4cm}  (\nu = 2 , 3).
   \end{cases} \label{eq:A_9}
\end{equation}
From Eq. (\ref{eq:A_6}), we obtain the coefficient $A^{\nu \nu'}_{\bm{k}, sA, s'A} (\bm{0})$ for $s = s'$,
\begin{equation}
A^{\nu \nu'}_{\bm{k}, sA, s'A} (\bm{0}) =
   \begin{cases}
     \frac{1}{4}  T^{4}_{\bm{k}}                      \hspace{1.825cm}  (\omega_{2} = \omega^{'}_{2} = 1),  \\
     \\
     \frac{1}{4} (1 - T^{2}_{\bm{k}})^{2}              \hspace{0.9cm}    (\omega_{2} = \omega^{'}_{2} = 2),  \\
     \\
     \frac{1}{4}  T^{2}_{\bm{k}} (1 - T^{2}_{\bm{k}})    \hspace{0.67cm}    (\omega_{2} \ne \omega^{'}_{2}),
   \end{cases} \label{eq:A_10}
\end{equation}
and for $s \ne s'$,
\begin{equation}
A^{\nu \nu'}_{\bm{k}, sA, s'A} (\bm{0}) =
   \begin{cases}
     (-1)^{\nu + \nu'} \frac{1}{4}  T^{4}_{\bm{k}}                      \hspace{1.825cm}  (\omega_{2} = \omega^{'}_{2} = 1),  \\
     \\
     (-1)^{\nu + \nu'} \frac{1}{4} (1 - T^{2}_{\bm{k}})^{2}              \hspace{0.9cm}    (\omega_{2} = \omega^{'}_{2} = 2),  \\
     \\
     (-1)^{\nu + \nu'} \frac{1}{4}  T^{2}_{\bm{k}} (1 - T^{2}_{\bm{k}})    \hspace{0.67cm}    (\omega_{2} \ne \omega^{'}_{2}).
   \end{cases} \label{eq:A_11}
\end{equation}
Similarly, $A^{\nu \nu'}_{\bm{k}, sA, s'B} (\bm{0})$ is obtained as follows. 
For $s = s'$,
\begin{equation}
A^{\nu \nu'}_{\bm{k}, sA, s'B} (\bm{0}) =
   \begin{cases}
     \frac{1}{4}  T^{2}_{\bm{k}} (1 - T^{2}_{\bm{k}})                 \hspace{1.48cm}  (\omega_{1} = \omega^{'}_{1}),  \\
     \\
     -\frac{1}{4}  T^{2}_{\bm{k}} (1 - T^{2}_{\bm{k}})                   \hspace{1.27cm}  (\omega_{1} \ne \omega^{'}_{1}).
   \end{cases} \label{eq:A_12}
\end{equation}
For $s \ne s'$,
\begin{equation}
A^{\nu \nu'}_{\bm{k}, sA, s'B} (\bm{0}) =
   \begin{cases}
     \frac{1}{4}  T^{2}_{\bm{k}} (1 - T^{2}_{\bm{k}})                 \hspace{1.48cm}  (\omega_{2} = \omega^{'}_{2}),  \\
     \\
     -\frac{1}{4}  T^{2}_{\bm{k}} (1 - T^{2}_{\bm{k}})                   \hspace{1.27cm}  (\omega_{2} \ne \omega^{'}_{2}).
   \end{cases} \label{eq:A_13}
\end{equation}

Now, we recast $A^{\nu \nu'}_{\bm{k}, sl, s'l'} (\bm{0})$ as
\begin{equation}
A^{\nu \nu'}_{\bm{k}, sl, s'l'} (\bm{0}) = \eta^{\nu \nu'}_{ sl, s'l'} S^{\nu \nu'}_{\bm{k},l l'}, \label{eq:A_14}
\end{equation} 
where $\eta^{\nu \nu'}_{ sl, s'l'}$ takes $+1$ or $-1$. 
It has been shown that 
$S^{\nu \nu'}_{\bm{k},AA}$ is one of the following $S_{\bm{k} i}$ ($i = 1, 2, 3$); 
\begin{align}
S_{\bm{k} 1} &= \frac{1}{4}  T^{4}_{\bm{k}}                    , \notag \\
S_{\bm{k} 2} &= \frac{1}{4} (1 - T^{2}_{\bm{k}})^{2}            , \notag \\
S_{\bm{k} 3} &= \frac{1}{4}  T^{2}_{\bm{k}} (1 - T^{2}_{\bm{k}})  . \label{eq:A_15}
\end{align} 
On the other hand, we obtain $S^{\nu \nu'}_{\bm{k},AB} = S_{\bm{k} 3}$.
We summarize $\eta^{\nu \nu'}_{ sl, s'l'}$ and $S^{\nu \nu'}_{\bm{k}, l l'}$ in Tables~\ref{Table:1} and \ref{Table:2}.
In order to calculate the irreducible multipole susceptibility [Eq.~(\ref{eq:21})], the summation with respect to 
the spin indices $s$ and $s'$ should be carried out. 
Thus, we also show $\sum_{s,s'} \eta^{\nu \nu'}_{s l, s' l'}$ in Tables \ref{Table:1} and \ref{Table:2}.

\begin{table}[!htb]
  \centering
  %\scalebox{1.0}{
  \begin{tabular}{|c|ccccc|c|}
    \hline
    \parbox[c][0.5cm][c]{0cm}{} $\nu \nu'$  & $\eta^{\nu \nu'}_{\uparrow A, \uparrow A}$ & $\eta^{\nu \nu'}_{\uparrow A, \downarrow A}$ & $\eta^{\nu \nu'}_{\downarrow A, \uparrow A}$ & $\eta^{\nu \nu'}_{\downarrow A, \downarrow A}$ & $\sum_{s,s'} \eta^{\nu \nu'}_{s A, s' A}$ & $S^{\nu \nu'}_{\bm{k}, AA}$    \\
    \hline
    \parbox[c][0.4cm][c]{0cm}{}         11    & $+1$ & $+1$ & $+1$ & $+1$ & $+4$ & $S_{\bm{k} 1}$ \\
    \parbox[c][0.4cm][c]{0cm}{}         12    & $+1$ & $-1$ & $-1$ & $+1$ & $0$  & $S_{\bm{k} 3}$ \\
    \parbox[c][0.4cm][c]{0cm}{}         13    & $+1$ & $+1$ & $+1$ & $+1$ & $+4$ & $S_{\bm{k} 3}$ \\
    \parbox[c][0.4cm][c]{0cm}{}         14    & $+1$ & $-1$ & $-1$ & $+1$ & $0$  & $S_{\bm{k} 1}$ \\
                                              &      &      &      &      &      &             \\
    \parbox[c][0.4cm][c]{0cm}{}         21    & $+1$ & $-1$ & $-1$ & $+1$ & $0$  & $S_{\bm{k} 3}$ \\
    \parbox[c][0.4cm][c]{0cm}{}         22    & $+1$ & $+1$ & $+1$ & $+1$ & $+4$ & $S_{\bm{k} 2}$ \\
    \parbox[c][0.4cm][c]{0cm}{}         23    & $+1$ & $-1$ & $-1$ & $+1$ & $0$  & $S_{\bm{k} 2}$ \\
    \parbox[c][0.4cm][c]{0cm}{}         24    & $+1$ & $+1$ & $+1$ & $+1$ & $+4$ & $S_{\bm{k} 3}$ \\
                                              &      &      &      &      &      &             \\
    \parbox[c][0.4cm][c]{0cm}{}         31    & $+1$ & $+1$ & $+1$ & $+1$ & $+4$ & $S_{\bm{k} 3}$ \\
    \parbox[c][0.4cm][c]{0cm}{}         32    & $+1$ & $-1$ & $-1$ & $+1$ & $0$  & $S_{\bm{k} 2}$ \\
    \parbox[c][0.4cm][c]{0cm}{}         33    & $+1$ & $+1$ & $+1$ & $+1$ & $+4$ & $S_{\bm{k} 2}$ \\
    \parbox[c][0.4cm][c]{0cm}{}         34    & $+1$ & $-1$ & $-1$ & $+1$ & $0$  & $S_{\bm{k} 3}$ \\
                                              &      &      &      &      &      &             \\
    \parbox[c][0.4cm][c]{0cm}{}         41    & $+1$ & $-1$ & $-1$ & $+1$ & $0$  & $S_{\bm{k} 1}$ \\
    \parbox[c][0.4cm][c]{0cm}{}         42    & $+1$ & $+1$ & $+1$ & $+1$ & $+4$ & $S_{\bm{k} 3}$ \\
    \parbox[c][0.4cm][c]{0cm}{}         43    & $+1$ & $-1$ & $-1$ & $+1$ & $0$  & $S_{\bm{k} 3}$ \\
    \parbox[c][0.4cm][c]{0cm}{}         44    & $+1$ & $+1$ & $+1$ & $+1$ & $+4$ & $S_{\bm{k} 1}$ \\
    \hline
  \end{tabular}
  %}
  \caption{Table of $\eta^{\nu \nu'}_{ sA, s'A}$,  $\sum_{s,s'} \eta^{\nu \nu'}_{s A, s' A}$, and $S^{\nu \nu'}_{\bm{k},AA}$. 
  }
  \label{Table:1}
\end{table}

\begin{table}[!htb]
  \centering
  %\scalebox{1.0}{
  \begin{tabular}{|c|ccccc|c|}
    \hline
    \parbox[c][0.5cm][c]{0cm}{} $\nu \nu'$  & $\eta^{\nu \nu'}_{\uparrow A, \uparrow B}$ & $\eta^{\nu \nu'}_{\uparrow A, \downarrow B}$ & $\eta^{\nu \nu'}_{\downarrow A, \uparrow B}$ & $\eta^{\nu \nu'}_{\downarrow A, \downarrow B}$ & $\sum_{s,s'} \eta^{\nu \nu'}_{s A, s' B}$ & $S^{\nu \nu'}_{\bm{k}, AB}$    \\
    \hline
    \parbox[c][0.4cm][c]{0cm}{}         11    & $+1$ & $+1$ & $+1$ & $+1$ & $+4$ & $S_{\bm{k} 3}$ \\
    \parbox[c][0.4cm][c]{0cm}{}         12    & $+1$ & $-1$ & $-1$ & $+1$ & $0$  & $S_{\bm{k} 3}$ \\
    \parbox[c][0.4cm][c]{0cm}{}         13    & $-1$ & $-1$ & $-1$ & $-1$ & $-4$ & $S_{\bm{k} 3}$ \\
    \parbox[c][0.4cm][c]{0cm}{}         14    & $-1$ & $+1$ & $+1$ & $-1$ & $0$  & $S_{\bm{k} 3}$ \\
                                              &      &      &      &      &      &             \\
    \parbox[c][0.4cm][c]{0cm}{}         21    & $+1$ & $-1$ & $-1$ & $+1$ & $0$  & $S_{\bm{k} 3}$ \\
    \parbox[c][0.4cm][c]{0cm}{}         22    & $+1$ & $+1$ & $+1$ & $+1$ & $+4$ & $S_{\bm{k} 3}$ \\
    \parbox[c][0.4cm][c]{0cm}{}         23    & $-1$ & $+1$ & $+1$ & $-1$ & $0$  & $S_{\bm{k} 3}$ \\
    \parbox[c][0.4cm][c]{0cm}{}         24    & $-1$ & $-1$ & $-1$ & $-1$ & $-4$ & $S_{\bm{k} 3}$ \\
                                              &      &      &      &      &      &             \\
    \parbox[c][0.4cm][c]{0cm}{}         31    & $-1$ & $-1$ & $-1$ & $-1$ & $-4$ & $S_{\bm{k} 3}$ \\
    \parbox[c][0.4cm][c]{0cm}{}         32    & $-1$ & $+1$ & $+1$ & $-1$ & $0$  & $S_{\bm{k} 3}$ \\
    \parbox[c][0.4cm][c]{0cm}{}         33    & $+1$ & $+1$ & $+1$ & $+1$ & $+4$ & $S_{\bm{k} 3}$ \\
    \parbox[c][0.4cm][c]{0cm}{}         34    & $+1$ & $-1$ & $-1$ & $+1$ & $0$  & $S_{\bm{k} 3}$ \\
                                              &      &      &      &      &      &             \\
    \parbox[c][0.4cm][c]{0cm}{}         41    & $-1$ & $+1$ & $+1$ & $-1$ & $0$  & $S_{\bm{k} 3}$ \\
    \parbox[c][0.4cm][c]{0cm}{}         42    & $-1$ & $-1$ & $-1$ & $-1$ & $-4$ & $S_{\bm{k} 3}$ \\
    \parbox[c][0.4cm][c]{0cm}{}         43    & $+1$ & $-1$ & $-1$ & $+1$ & $0$  & $S_{\bm{k} 3}$ \\
    \parbox[c][0.4cm][c]{0cm}{}         44    & $+1$ & $+1$ & $+1$ & $+1$ & $+4$ & $S_{\bm{k} 3}$ \\
    \hline
  \end{tabular}
  %}
  \caption{Table of $\eta^{\nu \nu'}_{ sA, s'B}$,  $\sum_{s,s'} \eta^{\nu \nu'}_{s A, s' B}$, and $S^{\nu \nu'}_{\bm{k}, AB}$. 
  }
  \label{Table:2}
\end{table}
%%%%%%%%%%%%%%%%%%%%%%%%%%%%%%%%%%%%%%%%%%%%%%%%%%%%%%%%%%%%%%%%%%%
%%%%%%%%%%%%%%%%%%%%%%%%%%%%%%%%%%%%%%%%%%%%%%%%%%%%%%%%%%%%%%%%%%%
%%%%%%%%%%%%%%%%%%%%%%%%%%%%%%%%%%%%%%%%%%%%%%%%%%%%%%%%%%%%%%%%%%%
%%%%%%%%%%%%%%%%%%%%%%%%%%%%%%%%%%%%%%%%%%%%%%%%%%%%%%%%%%%%%%%%%%%
%%%%%%%%%%%%%%%%%%%%%%%%%%%%%%%%%%%%%%%%%%%%%%%%%%%%%%%%%%%%%%%%%%%

By using the above results and the fact that $E_{\bm{k}1} = E_{\bm{k}2} \ne E_{\bm{k}3} = E_{\bm{k}4}$, 
we obtain the analytic forms of the irreducible susceptibility, Eqs.~(\ref{eq:27}) and (\ref{eq:28}).

\section{Emergent ASOC in EO state}

In this Appendix, we present the analytic calculation of the effective g-vector defined in Eq.~(\ref{eq:49}).
We begin by calculating the unitary matrix $\hat{U} (\bm{k})$, diagonalizing the mean field Hamiltonian 
$\hat{H}^{\, \rm{MF}}_{4} (\bm{k})$ [Eq.~(\ref{eq:11})], 
\begin{align}
\hat{U} & (\bm{k}) = \frac{1}{\sqrt{2}} \times \notag \\
&
\scalebox{1.025}{$\displaystyle
{\footnotesize
\begin{pmatrix}
T_{\bm{k -}}  &  -\frac{\lambda_{\bm{k}}^{+}}{|\bm{g}_{\bm{k}}|} \sqrt{1 - T_{\bm{k +}}^{2}}  &  \sqrt{1 - T_{\bm{k} -}^{2}}  &  -\frac{\lambda_{\bm{k}}^{+}}{|\bm{g}_{\bm{k}}|} T_{\bm{k} +} \\
\\
\frac{\lambda_{\bm{k}}^{-}}{|\bm{g}_{\bm{k}}|} T_{\bm{k} -}  &  \sqrt{1 - T_{\bm{k} +}^{2}}  &  \frac{\lambda_{\bm{k}}^{-}}{|\bm{g}_{\bm{k}}|} \sqrt{1 - T_{\bm{k} -}^{2}}  &  T_{\bm{k} +} \\
\\
\sqrt{1 - T_{\bm{k} -}^{2}}  &  -\frac{\lambda_{\bm{k}}^{+}}{|\bm{g}_{\bm{k}}|} T_{\bm{k} +}  &  -T_{\bm{k} -}  &  \frac{\lambda_{\bm{k}}^{+}}{|\bm{g}_{\bm{k}}|}  \sqrt{1 - T_{\bm{k} +}^{2}} \\
\\
\frac{\lambda_{\bm{k}}^{-}}{|\bm{g}_{\bm{k}}|}  \sqrt{1 - T_{\bm{k} -}^{2}}  &  T_{\bm{k} +}  &  -\frac{\lambda_{\bm{k}}^{-}}{|\bm{g}_{\bm{k}}|} T_{\bm{k} -}  &  -\sqrt{1 - T_{\bm{k} +}^{2}}
\end{pmatrix} ,
} $}
\label{eq:B_1}
\end{align}
where $T_{\bm{k} \pm}$ is defined by 
\begin{equation}
T_{\bm{k} \pm} \equiv \frac{t_{\perp}}{\sqrt{t^{2}_{\perp} + \left[\alpha^{\, d}_{\bm{k} \pm} + \sqrt{ (\alpha^{\, d}_{\bm{k} \pm})^{2} + t^{2}_{\perp}}\right]^{2}}}, \label{eq:B_2}
\end{equation}
and $\alpha^{\, d}_{\bm{k} \pm}$ is given by Eq.~(\ref{eq:54}). 
The dispersion relation of the $\nu$-th eigenstate is described in Eqs.~(\ref{eq:42a})-(\ref{eq:42d}). 
%\begin{align}
%E_{\bm{k} 1} &=  \varepsilon_{\bm{k}} + d_{\bm{k}} \frac{\Delta_{A} + \Delta_{B}}{2} + \sqrt{(\alpha^{\, d}_{\bm{k} -})^{2} + t^{2}_{\perp}} , \label{eq:B_5} \\
%E_{\bm{k} 2} &=  \varepsilon_{\bm{k}} + d_{\bm{k}} \frac{\Delta_{A} + \Delta_{B}}{2} + \sqrt{(\alpha^{\, d}_{\bm{k} +})^{2} + t^{2}_{\perp}} , \label{eq:B_6} \\
%E_{\bm{k} 3} &=  \varepsilon_{\bm{k}} + d_{\bm{k}} \frac{\Delta_{A} + \Delta_{B}}{2} - \sqrt{(\alpha^{\, d}_{\bm{k} -})^{2} + t^{2}_{\perp}} , \label{eq:B_7} \\
%E_{\bm{k} 4} &=  \varepsilon_{\bm{k}} + d_{\bm{k}} \frac{\Delta_{A} + \Delta_{B}}{2} - \sqrt{(\alpha^{\, d}_{\bm{k} +})^{2} + t^{2}_{\perp}} , \label{eq:B_8}
%\end{align}
%and summarized in Eq.~(\ref{eq:42}).

The expectation value of the spin operator $\bm{S}^{\, \rm{av}}_{\bm{k},\nu}$ is calculated 
with the use the unitary matrix $\hat{U} (\bm{k})$.
When we adopt the basis $\hat{C}^{\dagger}_{\bm{k}} = (c^{\dagger}_{\bm{k} \uparrow A}, c^{\dagger}_{\bm{k} \downarrow A}, c^{\dagger}_{\bm{k} \uparrow B}, c^{\dagger}_{\bm{k} \downarrow B})$, the spin operator is represented by the following $4 \times 4$ matrix;
\begin{equation}
\hat{S}^{x}_{4} =
\scalebox{1.0}{$\displaystyle
{\footnotesize
\begin{pmatrix}
0     & 1     & 0     & 0      \\
1     & 0     & 0     & 0      \\
0     & 0     & 0     & 1      \\
0     & 0     & 1     & 0
\end{pmatrix}
} $},
\label{eq:B_9}
\end{equation}

\begin{equation}
\hat{S}^{y}_{4}  =
\scalebox{1.0}{$\displaystyle
{\footnotesize
\begin{pmatrix}
0     & -i    & 0     & 0      \\
i     & 0     & 0     & 0      \\
0     & 0     & 0     & -i     \\
0     & 0     & i     & 0
\end{pmatrix}
} $},
\label{eq:B_10}
\end{equation}

\begin{equation}
\hat{S}^{z}_{4} =
\scalebox{1.0}{$\displaystyle
{\footnotesize
\begin{pmatrix}
1     & 0    & 0     & 0      \\
0     & -1   & 0     & 0      \\
0     & 0    & 1     & 0      \\
0     & 0    & 0     & -1
\end{pmatrix}
} $}.
\label{eq:B_11}
\end{equation}
As a result of the unitary transformation, the spin operator is transformed into 
\begin{align}
\hat{\bm{S}}_{\bm{k}} &= \hat{C}^{\dagger}_{\bm{k}} \, \hat{\bm{S}}_{4} \, \hat{C}_{\bm{k}} \notag \\
                  &= \underbrace{\hat{C}^{\dagger}_{\bm{k}} \hat{U}(\bm{k})}_{\hat{\Gamma}^{\dagger}_{\bm{k}}}  \underbrace{\hat{U}^{\dagger}(\bm{k}) \hat{\bm{S}}_{4} \hat{U}(\bm{k})}_{\tilde{\bm{S}}_{4} (\bm{k})}   \underbrace{\hat{U}^{\dagger}(\bm{k}) \hat{C}_{\bm{k}}}_{\hat{\Gamma}_{\bm{k}}} \notag \\
                  &= \hat{\Gamma}^{\dagger}_{\bm{k}} \, \tilde{\bm{S}}_{4} (\bm{k}) \, \hat{\Gamma}_{\bm{k}}, \label{eq:B_12}
\end{align}
where the band basis is represented by 
$\hat{\Gamma}^{\dagger}_{\bm{k}} = (\gamma^{\dagger}_{\bm{k} 1}, \gamma^{\dagger}_{\bm{k} 2}, \gamma^{\dagger}_{\bm{k} 3}, \gamma^{\dagger}_{\bm{k} 4})$. 
The diagonal matrix element of $\tilde{\bm{S}}_{4} (\bm{k})$ is nothing but the expectation value of the spin operator 
for each eigenstate.
Thus, we obtain 
\begin{align}
[\tilde{S}^{x}_{4} (\bm{k})]_{11} &=  [\tilde{S}^{x}_{4} (\bm{k})]_{33} = \frac{\sin k_{y}}{|\bm{g}_{\bm{k}}|}, \label{eq:B_13}  
\\
[\tilde{S}^{x}_{4} (\bm{k})]_{22} &= [\tilde{S}^{x}_{4} (\bm{k})]_{44}  = -\frac{\sin k_{y}}{|\bm{g}_{\bm{k}}|}, \label{eq:B_14}  
\end{align} 
and
\begin{align}
[\tilde{S}^{y}_{4} (\bm{k})]_{11} &= [\tilde{S}^{y}_{4} (\bm{k})]_{33}  = -\frac{\sin k_{x}}{|\bm{g}_{\bm{k}}|}, \label{eq:B_17} 
\\
[\tilde{S}^{y}_{4} (\bm{k})]_{22} &= [\tilde{S}^{y}_{4} (\bm{k})]_{44}  =  \frac{\sin k_{x}}{|\bm{g}_{\bm{k}}|}. \label{eq:B_18} 
\end{align}
We confirmed that 
\begin{align}
[\tilde{S}^{z}_{4} (\bm{k})]_{11} &= [\tilde{S}^{z}_{4} (\bm{k})]_{22} 
= [\tilde{S}^{z}_{4} (\bm{k})]_{33} = [\tilde{S}^{z}_{4} (\bm{k})]_{44} =0.
\label{eq:B_19}
\end{align}
From Eqs.~(\ref{eq:B_13})-(\ref{eq:B_19}), the expectation value is represented as
\begin{align}
\bm{S}^{\, \rm{av}}_{\bm{k},1} &= \bm{S}^{\, \rm{av}}_{\bm{k},3} = \frac{1}{\sqrt{ \sin^{2} k_{x} + \sin^{2} k_{y}} } \hspace{1mm} \biggl(\sin k_{y} ,\hspace{0.5mm} -\sin k_{x} ,\hspace{0.5mm} 0\biggr), \label{eq:B_25} \\
\bm{S}^{\, \rm{av}}_{\bm{k},2} &= \bm{S}^{\, \rm{av}}_{\bm{k},4} = \frac{1}{\sqrt{ \sin^{2} k_{x} + \sin^{2} k_{y}} } \hspace{1mm} \biggl(-\sin k_{y} ,\hspace{0.5mm} \sin k_{x} ,\hspace{0.5mm} 0\biggr). \label{eq:B_26}
\end{align}
By using Eqs.~(\ref{eq:42a})-(\ref{eq:42d}), (\ref{eq:B_25}), and (\ref{eq:B_26}), 
the effective g-vector is obtained as Eqs.~(\ref{eq:52}) and (\ref{eq:53}).

%%%%%%%%%%%%%%%%%%%%%%%%%%%%%%%%%%%%%%%%%%%%%%%%%%%%%%%%%%%%%%%%%%%
%%%%%%%%%%%%%%%%%%%%%%%%%%%%%%%%%%%%%%%%%%%%%%%%%%%%%%%%%%%%%%%%%%%
%%%%%%%%%%%%%%%%%%%%%%%%%%%%%%%%%%%%%%%%%%%%%%%%%%%%%%%%%%%%%%%%%%%
%%%%%%%%%%%%%%%%%%%%%%%%%%%%%%%%%%%%%%%%%%%%%%%%%%%%%%%%%%%%%%%%%%%
%%%%%%%%%%%%%%%%%%%%%%%%%%%%%%%%%%%%%%%%%%%%%%%%%%%%%%%%%%%%%%%%%%%

%%%%%%%%Bibiography Style File for JPSJ %%%%%%%%%%%%%
% Released on November 15, 1996: Version 1.00       %
% Copyright (C) 1996 by Shinsaku Fujita,            %
%                             all rights reserved.  %
%%%%%%%%%%Bibliography%%%%%%%%%%%%%%%%%%%%%%%%%%%%%%%

% Produces the bibliography via BibTeX.

\begin{thebibliography}{10}
\bibitem{Comment1}
In three dimensions, the g-vector with $k_{y} \hat{x} + k_{x} \hat{y} + \beta k_x k_y k_z \hat{z}$ symmetry is allowed. 
Since we study the 2D system, the last term disappears.
\bibitem{Comment2}
The $C_4$ rotation symmetry is slightly broken in YBCO owing to CuO chains. 
Thus, a weak nematicity is observed in the entire temperature range.
\end{thebibliography}
\end{document}